\documentclass[aps,prd,twocolumn,floatfix,showpacs,superscriptaddress,nofootinbib]{revtex4-2}
\usepackage{amsmath,amsfonts,amssymb,bm}
\usepackage{graphicx}
\usepackage{color}
\usepackage{subfigure}
\usepackage{multirow}
\usepackage{textcomp}
\usepackage{slashed}
\usepackage{bbm}
\usepackage{mathrsfs}
\usepackage[bottom]{footmisc}
\definecolor{purple}{rgb}{0.5,0,0.5}
\definecolor{blue}{rgb}{0.0,0,1.0}
\usepackage[colorlinks=true, pdfstartview=FitV, linkcolor=purple, citecolor= purple, urlcolor=blue]{hyperref}

\begin{document}
\title{Mapping spatial distributions within pseudoscalar mesons}

\author{Kh\'epani Raya}%
\email{khepani.raya@dci.uhu.es}
\affiliation{Department of Integrated Sciences and Center for Advanced Studies in Physics, Mathematics and Computation, University of Huelva, E-21071 Huelva, Spain}

\author{Adnan Bashir}
\email{adnan.bashir@dci.uhu.es}
\affiliation{Department of Integrated Sciences and Center for Advanced Studies in Physics, Mathematics and Computation, University of Huelva, E-21071 Huelva, Spain}
\affiliation{Instituto de F\'isica y Matem\'aticas, Universidad Michoacana de San Nicol\'as de Hidalgo, Morelia, Michoac\'an 58040, Mexico}

\author{Jos\'e Rodr\'iguez-Quintero}
\email{jose.rodriguez@dfaie.uhu.es}
\affiliation{Department of Integrated Sciences and Center for Advanced Studies in Physics, Mathematics and Computation, University of Huelva, E-21071 Huelva, Spain}
\affiliation{Irfu, CEA, Université de Paris-Saclay, 91191, Gif-sur-Yvette, France}

\date{\today}
\begin{abstract}
Several aspects of the internal structure of pseudoscalar mesons, accessible through generalized parton distributions in their zero-skewness limit, are examined. These include electromagnetic and gravitational form factors related to charge and mass densities; and distributions in the impact parameter space. To this end, we employ an  algebraically viable framework that is based upon the valence-quark generalized parton distribution expressed explicitly in terms of the associated distribution function and a profile function that governs the off-forward dynamics. The predominantly analytical nature of this scheme  yields several algebraic results and relations while also facilitating the exploration of insightful limiting cases. With a suitable input distribution function, guided either by experiment or theory, and with an appropriate choice of the profile function, it is possible to provide testable predictions for spatial distributions of valence quarks inside pseudoscalar mesons. When comparison is possible, these predictions align well with existing experimental data as well as the findings of reliable theoretical approaches and lattice QCD.
\end{abstract}

\maketitle

\section{Introduction}

Despite the successful history of quantum chromodynamics (QCD),\,\cite{Gross:2022hyw}, spanning over half a century, its unparalleled emergent features not only continue to boggle the mind but also demand vigorous research
in understanding hadron structure and properties.
 Firstly, the fundamental degrees of freedom—namely, quarks and gluons—cannot be detected in isolation; rather, due to color confinement, the observable entities are color-singlet bound-states, \emph{i.e.}, hadrons. Secondly, the percentage difference between the proton mass and its composing current quarks is by far the largest as compared to any other stable composite system. These uncanny features can be attributed to the fact that while asymptotic freedom results in a reduction of the strong interaction coupling at high energy scales, enabling a perturbative treatment of QCD, many static and dynamic characteristics of hadrons (including their masses) are driven by non-perturbative phenomena\,\cite{Roberts:2020hiw}. 
 Traditionally, the internal structure of hadrons has been explored mainly through the use of electromagnetic probes, revealing, for example, the proton's non-point-like nature, leading to the subsequent discovery of quarks~\cite{Ellis:1996mzs,Hofstadter:1956qs,Breidenbach:1969kd}. The experimental principles remain in force even though technology has improved substantially\,\cite{Accardi:2023chb, Anderle:2021wcy, Accardi:2012qut, Quintans:2022utc}. Modern experiments involving exclusive reactions offer access to  electromagnetic form factors (EFFs) and linked distribution amplitudes (DAs) in the large momentum transfer regime~\cite{Lepage:1980fj}, while the so called parton densities and distribution functions (DFs) are usually inferred from inclusive processes~\cite{Ellis:1996mzs}. Generalized parton distributions (GPDs), introduced 
about three decades ago~\cite{Muller:1994ses,Ji:1996nm,Ji:1996ek,Radyushkin:1996ru,Radyushkin:1997ki}, unify all of those objects and provide a three-dimensional picture of hadrons (often referred to as hadron tomography) as well as a connection with the energy momentum tensor~\cite{Polyakov:2002yz,Polyakov:2018zvc}. Defined in the impact parameter space (IPS), GPDs provide a spatial visualization of parton distributions~\cite{Diehl:2002he,Burkardt:2000za}, thus completing a detailed picture of the internal structure of hadrons.

The empirical extraction of GPDs is possible through exclusive processes~\cite{Diehl:2003ny,Belitsky:2005qn,Guidal:2013rya,Qiu:2022bpq,Qiu:2022pla,Qiu:2023mrm}, but it presents a multitude of challenges, in particular its $x$-dependence. The main reason lies within the indirect nature of their identification, which requires specific kinematic conditions to be met and a careful interpretation of the measured cross sections\,\cite{Mezrag:2023nkp,Chavez:2021koz}. Concerning this matter, methods have been proposed to infer the GPDs from limited knowledge of specific kinematic regions\,\cite{Chouika:2017dhe,Chouika:2017rzs,Chavez:2021llq,DallOlio:2024vjv}, even suggesting that they can be reconstructed through prior determination of seemingly simpler distributions, such as DFs and EFFs\,\cite{Xu:2023bwv}. From a theoretical standpoint the challenge is by no means minor, since a rigorous construction of such requires a number of mathematical and physical constraints to be fulfilled\,\cite{Mezrag:2023nkp}. Nonetheless, significant progress has been made through lattice and continuum QCD methods, \emph{e.g.}\,\cite{Riberdy:2023awf,Constantinou:2020hdm,Lin:2023gxz,Roberts:2021nhw,Ding:2022ows,Raya:2024ejx}. While the proton has traditionally been the primary focus of study, it is becoming increasingly evident that there is a pressing need to investigate the internal structure of both the pion and the kaon with equal rigor\,\cite{Aguilar:2019teb,Arrington:2021biu}. Firstly, these bound-states act as conveyors of the strong force in terms of hadron degrees of freedom; secondly, their identity as Nambu-Goldstone (NG) bosons of dynamical chiral symmetry breaking, closely ties them to the processes underlying the mass generation of visible matter\,\cite{Horn:2016rip,Roberts:2020hiw}. A full picture would be incomplete without a comparison of their properties with those of the heavier pseudoscalar mesons, which are more strongly coupled with the Higgs mass generation\,\cite{Raya:2024ejx}. In light of the above, this work proposes a concise yet illustrative representation of the valence-quark GPD for pseudoscalar mesons. This characterization facilitates the direct derivation of various measurable quantities, establishing a connection between them. We refer to this framework as the Exponential Representation Scheme (ERS). With appropriate inputs, the ERS shows consistency with predictions derived from more sophisticated approaches, \emph{e.g.} Refs.\,\cite{Ding:2022ows,Raya:2024ejx}, while also aligning with experimental data and lattice QCD results when available\,\cite{Lu:2023yna,Cui:2021mom,Cui:2022bxn}. 

The manuscript is organized as follows: Section\,\ref{sec:ERS} outlines the ERS scheme,  while Section\,\ref{sec:facAns} covers general aspects of factorized models. Instructive special cases are also examined. Section\,\ref{sec:numericalRes} addresses the findings concerning pseudoscalar mesons, focusing on the $\pi,\,K,$ and $D$ mesons, as well as the illustrative  $\pi_s$ and $\pi_c$ systems (positively charged pseudoscalars with valence-quark masses equivalent to those of the $s$ and $c$-quarks, respectively). A brief summary is then provided in Section\,\ref{sec:summary}.

\section{The exponential representation scheme}
\label{sec:ERS}

Let $\textbf{P}=q\bar{h}$ be a pseudoscalar meson composed of a $q\,(\bar{h})$ flavored valence-quark (antiquark). For such a system, we advocate the following form of the valence-quark zero-skewness GPD:
\begin{equation}
\label{eq:GPDgen}
H_{\textbf{P}}^q(x,\Delta^2; \zeta)=q_{\textbf{P}}(x;\zeta)\,\text{exp}\left[-\Delta^2 \hat{\phi}_{\textbf{P}}(x ;\zeta)\right]\,,
\end{equation}
at a given resolving scale\footnote{The GPD \eqref{eq:GPDgen} can be evolved from the resolving scale $\zeta$ to any other $\zeta'$, but it cannot be generally guaranteed the existence of a new $\hat{\phi}_\textbf{P}(x;\zeta')$ such that the 
evolved GPD may still be accommodated into the same form \eqref{eq:GPDgen}.} $\zeta$ which, as will be discussed below, can be identified as the natural one for the representation.
The kinematic variables are defined as usual: $x$ denotes the average light-front longitudinal momentum fraction; $\Delta^2$ is the momentum transferred by the probe. It is important to note that throughout this manuscript, we consider the zero-skewness limit ($\xi=0$) of the GPD, restricting us to the DGLAP kinematic domain\,\cite{Dokshitzer:1977sg,Gribov:1971zn,Lipatov:1974qm,Altarelli:1977zs}.
It aligns with the goal of our current analysis, although extensions to the full GPD kinematic domain, based on the preservation of Lorentz covariance, are also possible\,\cite{Chouika:2017rzs,Chouika:2017dhe,Chavez:2021llq,Chavez:2021koz,DallOlio:2024vjv}.

Eq.\,\eqref{eq:GPDgen} encompasses some previous proposals to model GPD as will be shortly discussed. Notice that the two functions  $q_{\textbf{P}}(x;\zeta)$ and $\hat{\phi}_{\textbf{P}}(x;\zeta)$ completely parameterize the GPD. In the forward limit ($\Delta=0$), the GPD specializes to the valence-quark DF, denoted in Eq.\,\eqref{eq:GPDgen} as $q_{\textbf{P}}(x;\zeta)$. The second term, $\hat{\phi}_{\textbf{P}}(x;\zeta)$, drives the $\Delta^2$ evolution of the GPD. Depending on the form chosen for $\hat{\phi}_{\textbf{P}}(x)$, one can recover the representation from the light-front holographic QCD (LFHQCD) approach\,\cite{deTeramond:2018ecg,Chang:2020kjj,Wang:2024sqg,Wang:2024fjt}, the Gaussian model (GM) discussed in Refs.\,\cite{Raya:2021zrz,Zhang:2021mtn}, or can be specialized to different algebraic schemes based on integral representations of mesons' Bethe-Salpeter amplitudes (see, \emph{e.g.}, Refs.\,\cite{Raya:2021zrz,Albino:2022gzs,Almeida-Zamora:2023bqb}).  Moreover, as will be shown below, several insights and useful relationships can be derived without specifying either the DF or $\hat{\phi}_{\textbf{P}}(x;\zeta)$. Several of its implications are discussed below.

\subsection{Light-front wave functions}
Let us first consider the overlap representation of the GPD\,\cite{Diehl:2000xz,Burkardt:2002hr,Mezrag:2016hnp}:
\begin{equation}
H^q_{\textbf{P}}(x,\Delta^2;\zeta) = \int \frac{d^2k_\perp}{16 \pi^3} 
\psi^{q\ast}_{\textbf{P}}\left(x,\textbf{k}_{\perp+}^2;\zeta \right)  
\psi^{q}_{\textbf{P}}\left( x,\textbf{k}_{\perp-}^2;\zeta \right) \,.
\label{eq:overlap}
\end{equation} 
Here, $\psi^{q}_{\textbf{P}}\left( x,k_\perp;\zeta \right)$ is the leading-twist light-front wave function (LFWF), $\textbf{k}_{\perp \pm}=k_{\perp}\pm(1-x)\Delta_\perp/2$, and $\Delta_\perp=\Delta$ in the limit of zero-skewness.  Arguably the simplest LFWF connecting Eq. \eqref{eq:GPDgen} and Eq.\,\eqref{eq:overlap} is:
\begin{equation}
\label{eq:LFWFgen}
    \psi_\textbf{P}^q(x,k_\perp^2;\zeta)=8\pi \frac{\sqrt{q_\textbf{P}(x;\zeta)\hat{\phi}_\textbf{P}(x;\zeta)}}{1-x}\,\text{exp}\left[-2k_\perp^2 \frac{\hat{\phi}_\textbf{P}(x;\zeta)}{(1-x)^2} \right]\,.
\end{equation}
Using this representation, $\psi_\textbf{P}^q(x,k_\perp^2;\zeta)$ is formulated in terms of the DF, and the $x-k_\perp$ correlations are controlled  by the profile function $\hat{\phi}_\textbf{P}(x;\zeta)$.


The distribution amplitude (DA) associated with the LFWF, $\varphi_{\textbf{P}}^q(x;\zeta)$, is obtained by the simple integration:
\begin{equation}
\label{eq:PDAdef}
 f_{\textbf{P}}\varphi_{\textbf{P}}^q(x;\zeta)=\int \frac{d^2k_\perp}{16 \pi^3} \psi^{q}_{\textbf{P}}\left( x,k_\perp^2;\zeta \right) \,,
\end{equation}
where $f_{\textbf{P}}$ is the meson's leptonic decay constant. Within the ERS, the unit-normalized DA reads:
\begin{equation}
    \label{eq:PDAmodel}    \varphi_{\textbf{P}}^q(x;\zeta)=\frac{1}{4\pi f_{\textbf{P}}}(1-x)\sqrt{\frac{q_\textbf{P}(x;\zeta)}{\hat{\phi}_{\textbf{P}}(x;\zeta})}\,,
\end{equation}
establishing a closed relationship between DA and DF, which is left fully determined only  by the choice of the profile function ${\hat{\phi}}_{\textbf{P}}(x;\zeta)$.

\subsection{Impact parameter space GPD}
The impact parameter space (IPS) GPD is obtained by evaluating a two-dimensional Fourier transform\,\cite{Diehl:2002he,Burkardt:2000za}:
\begin{equation}
    q_\textbf{P}(x,|b_\perp|;\zeta)=\int_0^\infty \frac{d\Delta}{2\pi}\Delta J_0(|b_\perp|\Delta)\,H_\textbf{P}^q(x,\Delta^2;\zeta)\,,
\end{equation}
where $J_0$ is a cylindrical Bessel function. This density encodes the probability of finding the valence quark $q$ at a transverse distance $|b_\perp|$ from the meson's centre of transverse momentum. Given the GPD representation in Eq.\,\eqref{eq:GPDgen}, it is straightforward to derive analytic expression (here $\mathcal{I}_\textbf{P}^q(x,|b_\perp|;\zeta):=2\pi|b_\perp|q_\textbf{P}(x,|b_;\zeta)$):
\begin{equation}
    \label{eq:IPSdef}
    \mathcal{I}_\textbf{P}^q(x,|b_\perp|;\zeta)=\frac{q_\textbf{P}(x;\zeta)}{2\hat{\phi}_\textbf{P}(x;\zeta)}|b_\perp|\,\text{exp}\left[-\frac{|b_\perp|^2}{4\hat{\phi}_\textbf{P}(x;\zeta)}\right]\,,
\end{equation}
whose global maximum of $\mathcal{I}_\textbf{P}^q(x,|b_\perp|;\zeta)$ is located at:
\begin{eqnarray}
    |b_\perp^{\text{max}}(\zeta)|&=&+\sqrt{2\, \hat{\phi}_\textbf{P}(x^{\text{max}};\zeta)}\,,
\end{eqnarray}
such that $x^{\text{max}}$ is a real-valued solution of:
\begin{equation}
    q_\textbf{P}(x;\zeta)\hat{\phi}_\textbf{P}'(x;\zeta)-2q_\textbf{P}'(x;\zeta)\hat{\phi}_\textbf{P}(x;\zeta)=0\,.
\end{equation}
The maximum of this distribution relies on key aspects of the meson's internal dynamics, particularly those associated with dynamical mass generation\,\cite{Raya:2021zrz}.

One could also consider the mean-squared transverse extent (MSTE):
\begin{equation}
\label{eq:MSTE}
    \langle b_{\perp}^2(x;\zeta)\rangle_\textbf{P}^q:=\int_0^\infty |b_\perp|^2 \mathcal{I}_\textbf{P}^q(x,|b_\perp|;\zeta) = 4 \hat{\phi}_{\textbf{P}}(x;\zeta)q_{\textbf{P}}(x;\zeta)\;,
\end{equation}
that yields the expectation value
\begin{equation}
\label{eq:MSTEave}
    \langle |b_{\perp}(\zeta)|^2\rangle_\textbf{P}^q:=\int_0^1 \langle b_{\perp}^2(x;\zeta)\rangle_\textbf{P}^q\,=4\langle x^0 \rangle_{\hat{\phi}q}\,.
\end{equation}
Here $\langle x^n\rangle_{\hat{\phi}q}$ corresponds to the $n-$th Mellin moment of the distribution $[\hat{\phi}_{\textbf{P}}(x;\zeta)q_\textbf{P}(x;\zeta)]$. 

\subsection{Electromagnetic and gravitational form factors}
The contribution of the $q$ valence-quark to the meson's electromagnetic form factor (EFF), $F_{\textbf{P}}(\Delta^2)$, is expressed through the zeroth moment of the GPD\,\cite{Mezrag:2023nkp}:
\begin{equation}
\label{eq:EFFq}
    F_{\textbf{P}}^q(\Delta^2)=\int_0^1dx\,H_\textbf{P}^q(x,\Delta^2;\zeta)\,;
\end{equation}
and, the complete EFF results from adding up the individual components:
\begin{equation}
\label{eq:EFFtot}
    F_{\textbf{P}}(\Delta^2):=e_q F_{\textbf{P}}^q(\Delta^2)+e_{\bar{h}}F_{\textbf{P}}^{\bar{h}}(\Delta^2)\,,
\end{equation}
where $e_q$, $e_{\bar{h}}$ are the valence-constituent electric charges in units of the positron charge. The integration domain in Eq.\,\eqref{eq:EFFq} reflects our choice to adopt the zero-skewness limit, since the EFF is independent of this kinematic variable\,\cite{Mezrag:2023nkp}. In this case, the $q$ valence-quark contribution to the gravitational form factor $\theta_2^{\textbf{P}}(\Delta^2)$ is given by\,\cite{Polyakov:2018zvc}:
\begin{equation}
\label{eq:GFFq}
    \theta_2^{\textbf{P}q}(\Delta^2;\zeta)=\int_0^1dx\,x H_\textbf{P}^q(x,\Delta^2;\zeta)\,,
\end{equation}
in such a way that the total GFF is determined by summing over the different parton types $a$:
\begin{equation}
   \theta_2^{\textbf{P}}(\Delta^2):=\sum_{a} \theta_2^{\textbf{P}a}(\Delta^2;\zeta)\,.
\end{equation}
Note that, while the total form factors are scale invariant, the individual parton contributions to $\theta_2^{\textbf{P}}(\Delta^2)$ are, in fact, scale dependent. 

The valence-quark contributions, inferred from Eq.\,\eqref{eq:GFFq} with a GPD  expressed as in Eq.\,\eqref{eq:GPDgen}, are specific to the resolution scale 
$\zeta$ associated with this formulation.
 As discussed elsewhere\,\cite{Ding:2019lwe,Ding:2019qlr,Cui:2020tdf,Raya:2024ejx}, identifying this scale with the so-called hadron scale, $\zeta=\zeta_H$, is both convenient and intuitive, as it associates all hadron properties to its valence constituents alone. Consequently,
\begin{equation}
\label{eq:GFFtot}
   \theta_2^{\textbf{P}}(\Delta^2)=\theta_2^{\textbf{P}q}(\Delta^2;\zeta_H)+\theta_2^{\textbf{P}\bar{h}}(\Delta^2;\zeta_H)\,.
\end{equation}
 Using the so called all-orders  evolution approach, the contributions from different parton species at any other scale $\zeta > \zeta_H$ can be  derived\,\cite{Raya:2021zrz,Yao:2024ixu}.  
Concepts related to this scheme and the definition of a hadronic scale have been significantly refined in recent years and applied to various contexts involving DFs and GPDs, see \emph{e.g.}~Refs.\,\cite{Zhang:2021mtn,Cui:2020tdf,Raya:2021zrz,Xu:2023bwv,Lu:2023yna,Yin:2023dbw,Cui:2022bxn,Cui:2021mom,Lu:2022cjx,dePaula:2022pcb,Yao:2024ixu}. Regarding mesons, this framework explicitly links the valence-quark and antiquark DFs via momentum conservation:
\begin{equation}
    \label{eq:DFanti}
    \bar{h}_{\textbf{P}}(x;\zeta_H)=q_{\textbf{P}}(1-x;\zeta_H)\,.
\end{equation}
Therefore, to calculate all the antiquark-related quantities, it is sufficient to replace $q_\textbf{P}(x;\zeta_H) \to q_\textbf{P}(1-x;\zeta_H)$. In addition, in the isospin symmetric limit ($m_q=m_h$), 
\begin{equation}
    q_\textbf{P}(x;\zeta_H) \overset{m_q= m_h}{=}q_\textbf{P}(1-x;\zeta_H) \,,
\end{equation}
which results in a symmetric distribution at the hadronic scale. Naturally, the above expression would be valid in quarkonia systems, or in those where the isospin symmetry is reasonable, such as the pion. Several features of the symmetric DFs are discussed in Refs.\,\cite{Lu:2023yna,Cui:2022bxn}.

\subsection{Charge and mass radii}
\label{sec:ChargeMass}
The  charge radius $r_E^\textbf{P}$ is defined as usual:
\begin{eqnarray}
(r_E^{\textbf{P}})^2&=& e_q (r_E^{\textbf{P}q})^2+e_{\bar{h}}(r_E^{\textbf{P}\bar{h}})^2\,,\\
    (r_{E}^{\textbf{P}q})^2&:=&-6 \frac{\partial F_\textbf{P}^q(\Delta^2;\zeta_H)}{\partial \Delta^2} \Bigg|_{\Delta^2=0}\,.\label{eq:chargeAn}
\end{eqnarray}
The ERS permits the quark contribution to adopt the following form:
\begin{equation}
(r_E^{\textbf{P}q})^2=6\int_0^1 \hat{\phi}_{\textbf{P}}(x;\zeta_H)q_\textbf{P}(x;\zeta_H)=6\langle x^0\rangle_{\hat{\phi}q}\,.
\label{eq:radiiDEF}
\end{equation}
As far as $\theta_2^{\textbf{P}}(\Delta^2)$ is concerned, one can anagously define a mass radius $r_M^\textbf{P}$\,:
\begin{eqnarray}
    (r_M^{\textbf{P}})^2&=&  (r_M^{\textbf{P}q})^2+(r_M^{\textbf{P}\bar{h}})^2\,,\\
    (r_M^{\textbf{P}q})^2&=&6\int_0^1 x\hat{\phi}_{\textbf{P}}(x;\zeta_H)q_\textbf{P}(x;\zeta_H)=6\langle x\rangle_{\hat{\phi}q}\,.\label{eq:massAn}
\end{eqnarray}
Focusing on charged pseudoscalars, the ratio of the mass and charge spatial extents can be translated to\,:
\begin{equation}
\label{eq:masschargeQ}
     \mathcal{R}_0^\textbf{P}:=\left(\frac{r_M}{r_E} \right)^2=\frac{\langle x \rangle_{\hat{\phi}q}+\langle x \rangle_{\hat{\phi}\bar{h}}}{e_q\langle x^0 \rangle_{\hat{\phi}q}+e_{\bar{h}}\langle x^0 \rangle_{\hat{\phi}\bar{h}}}\,.
\end{equation}
In the isospin symmetric limit, $m_q=m_h$, this yields\,:
\begin{equation}
\label{eq:masschargeQiso}
     \mathcal{R}_0^\textbf{P}\overset{m_q=m_h} {=}\frac{2\langle x \rangle_{\hat{\phi}q}}{\langle x^0 \rangle_{\hat{\phi}q}}\,.
\end{equation}
More aspects related to these results will be discussed later. At this point, we turn our attention to the charge and mass distributions within the ERS.

\subsection{Charge and mass distributions}
The role of a valence-quark $q$ in the two-dimensional charge and mass distributions, $\rho_E^{\textbf{P}q}(b_\perp)$ and $\rho_M^{\textbf{P}q}(b_\perp)$, respectively, is encoded within the zeroth and first moments the IPS-GPD\,\cite{Raya:2022eqa,Xu:2023bwv}; that is:
\begin{eqnarray}
\label{eq:chargeMass}
    \rho_{\{E,M\}}^{\textbf{P}q}(b_\perp;\zeta)=\frac{1}{2\pi |b_\perp|}\int_0^{1}dx\, \{1,x\}\,\mathcal{I}_{\textbf{p}}^q(x,|b_\perp|;\zeta)\,.\hspace{0.5cm}
\end{eqnarray}
Setting $\zeta=\zeta_H$, the meson's total charge and mass distributions are:
\begin{eqnarray}\label{eq:chargeDist}
    \rho_{E}^{\textbf{P}}(b_\perp;\zeta_H)&=&e_q \rho_{E}^{\textbf{P}q}(b_\perp;\zeta_H)+e_{\bar{h}}\rho_{E}^{\textbf{P}\bar{h}}(b_\perp;\zeta_H)\,,\\
    \rho_{M}^{\textbf{P}}(b_\perp;\zeta_H)&=& \rho_{M}^{\textbf{P}q}(b_\perp;\zeta_H)+\rho_{M}^{\textbf{P}\bar{h}}(b_\perp;\zeta_H)\,.\label{eq:massDist}
\end{eqnarray}
As with the $F_\textbf{P}(\Delta^2)$ and $\theta_2^{\textbf{P}}(\Delta^2)$, the specific shape of the associated distributions strongly depends on the fine details of the DF and $\hat{\phi}_\textbf{P}(x;\zeta_H)$. Nonetheless, the simple form manifested by $\mathcal{I}_\textbf{P}(x,|b_\perp|;\zeta_H)$, Eq.\,\eqref{eq:IPSdef}, enables us to anticipate the low-$|b_\perp|$ behavior:
\begin{eqnarray}\nonumber
    \rho_{\{E,M\}}^{\textbf{P}q}(b_\perp;\zeta_H)&\overset{|b_\perp|\to 0}{\approx}&\frac{1}{2\pi}\int_0^1dx\,\{1,x\}\frac{\,q_\textbf{P}(x;\zeta_H)}{2\hat{\phi}_{\textbf{P}}(x;\zeta_H)}\\
   &\times&\left(1-\frac{b_\perp^2}{4\hat{\phi}_{\textbf{P}}(x;\zeta)}+\mathcal{O}(b_\perp^4)\right)\,.\label{eq:lowBrho}
\end{eqnarray}
In the following section, we present simplified parameterizations for the corresponding DFs and $\hat{\phi}_\textbf{P}(x;\zeta)$, allowing us to explore insightful limiting cases. Thereafter, more realistic representations will be studied for a variety of pseudoscalar mesons.

\section{Factorized Ansatze}
\label{sec:facAns}

It has been argued that a factorized form for the pseudoscalar meson leading-twist LFWF is reliable for integrated quantities\,\cite{Xu:2018eii}. In particular, it is shown to correctly describe the isospin-symmetric chiral limit\,\cite{Raya:2021zrz,Zhang:2021mtn}. That is, it is a fair approximation to express
\begin{equation}
\label{eq:PsiFac}
    \psi_{\textbf{P}}^q(x,k_\perp^2;\zeta_H)= \varphi_{\textbf{P}}^q(x;\zeta_H) \tilde{\psi}_{\textbf{P}}(k_\perp^2;\zeta_H)\,,
\end{equation}
where $\tilde{\psi}_{\textbf{P}}^q(k_\perp^2;\zeta_H)$ is a suitable chosen function. In combination with Eq.\,\eqref{eq:overlap}, the GPD derived from a factorized LFWF takes the following form\cite{Raya:2021zrz,Zhang:2021mtn}:
\begin{equation}
    H_\textbf{P}(x,\Delta^2;\zeta_H)=q_\textbf{P}(x;\zeta_H) \Phi_\textbf{P}(z_\Delta;\zeta_H)\,.
\end{equation}
Plainly, $\Phi_\textbf{P}(z;\zeta_H)$ controls the off-forward behavior of the GPD. It happens to depend on a single variable. In the zero-skewness limit, it reads as:  $z_\Delta:=\Delta^2(1-x)^2$. Moreover, specializing in the forward limit $\Delta^2=0$, and recalling Eq.\,\eqref{eq:PDAdef}, one obtains:
\begin{equation}
    \label{eq:facPDFPDA}
    q_\textbf{P}(x;\zeta_H)= n_\varphi [\varphi_{\textbf{P}}^q(x;\zeta_H)]^2\,,
\end{equation}
where $n_\phi$ ensures the unit-normalization of $q_\textbf{P}(x;\zeta_H)$. 

Correspondingly, in the ERS context, the IPS-GPD acquires the following general structure\,\cite{Xu:2023bwv}:
\begin{equation}
    \label{eq:facIPSGPD}
    q_\textbf{P}(x,|b_\perp|;\zeta_H)=\frac{q_{\textbf{P}}(x;\zeta_H)}{(1-x)^2} \Psi_{\textbf{P}}(z_b;\zeta_H)\,.
\end{equation}
In analogy with the GPD given by Eq.\,\eqref{eq:PsiFac},  
$\Psi_{\textbf{P}}(z_b;\zeta_H)$ expresses the dependence on $|b_\perp|$ through the single variable $z_b:=|b_\perp|/(1-x)$.

One of the most notable outcomes of the factorized representation is that it enables the mass-to-charge ratio, Eq.\,\eqref{eq:masschargeQiso}, to be expressed as follows\,\cite{Raya:2021zrz,Xu:2023bwv} in the isospin symmetry limit\,:
\begin{equation}
\label{eq:ratiofrac}
    \mathcal{R}^{\textbf{P}}_0\overset{\text{fac}}{:=}\frac{2\langle x^2(1-x)\rangle_{q_{\textbf{P}}}}{\langle x^2 \rangle_{q_{\textbf{P}}}}=\frac{\langle x(1-x)\rangle_{q_{\textbf{P}}}}{\langle x^2 \rangle_{q_{\textbf{P}}}}\,;
\end{equation}
that is, this quotient would be completely determined by the low-order Mellin moments of the DF. This result holds for the entire class of factorized models of pion-like systems\,\cite{Lu:2023yna,Cui:2022bxn}. We will return to this aspect later.

A series of studies on the pion and kaon,\,\emph{e.g.} Refs.\,\cite{Cui:2020tdf,Raya:2021zrz,Xu:2018eii,Xu:2023bwv}, have validated Eq.\,\eqref{eq:PsiFac} and Eq.\,\eqref{eq:facPDFPDA}. Furthermore, subsequent investigations from \emph{e.g.}\,\cite{Zhang:2020ecj,Albino:2022gzs,Almeida-Zamora:2023bqb,Serna:2024vpn} suggest that deviations from this factorization are on the order of $\mathcal{O}(\tilde{m}_\textbf{P}^2, \delta M_{\textbf{P}}^2)$, where
\begin{equation}
\label{eq:violation}
    \tilde{m}_\textbf{P}^2:=\frac{m_\textbf{P}^2}{(M_q+M_h)^2},\,\delta M_{\textbf{P}}^2:= \frac{|M_h^2-M_q^2|}{(M_q+M_h)^2}\,;
\end{equation}
here $m_\textbf{P}$ is the meson's mass and $M_{q,h}$ the constituent quark masses. For the pion, both quantities are entirely negligible, while for the kaon they remain markedly small. 

In light of these observations, we regard the factorized representation of the $\pi-K$ LFWFs as reliable. To take advantage of the ERS and its corollaries, and to be consistent with Eq.\,\eqref{eq:PDAmodel}, we require $\hat{\phi}_{\textbf{P}}(x;\zeta_H)\sim (1-x)^2$. Accordingly, the following form is proposed:
\begin{equation}
    \label{eq:fxprofile}
    \hat{\phi}_\textbf{P}(x;\zeta_H):=(1-x)^2/\Lambda_\textbf{P}^2\,,
\end{equation}
where $\Lambda_\textbf{P}$ is a mass scale to be established. The final element needed to fully determine the GPD is the corresponding DF. We will first examine some special cases which will lead us to a more general analysis afterwards.

\subsection{Flat-top DF}
Let us consider the parametric form of the GPD suggested in Eq.\,\eqref{eq:GPDgen}, together with the $\hat{\phi}_\textbf{P}(x;\zeta)$ profile from Eq.\,\eqref{eq:fxprofile}. We also assume a charged isospin-symmetric system, characterized by a constant DF,
\begin{equation}
    q_{ci}(x):=\Theta(x)\Theta(1-x)\,.
\end{equation}
Such a distribution is typical of the contact interaction (CI) model\,\cite{Roberts:2010rn,Xing:2023eed,Zhang:2020ecj}, and is self-consistent with a constant DA $\varphi_{ci}(x):=1$. This oversimplified distribution indicates that the valence-quark and antiquark share momentum equally across the full range of possible momentum fractions. The corresponding Mellin moments are:
\begin{equation}
    \label{eq:momsCI}
    \langle x^n \rangle_{q_{{ci}}}=\int_0^1 dx \, x^n q_{{ci}}(x) = \frac{1}{n+1}\,.
\end{equation}
A variety of closed-ended outcomes follow directly within this idealized scenario. First, the GPD has a trivial form:
\begin{equation}
    \label{eq:GPDSCI}
    H_{ci}^q(x,\Delta^2)=\Theta(x)\Theta(1-x)\,\text{exp}\left[-\frac{\Delta^2}{\Lambda_{ci}^2}(1-x)^2\right]\,.
\end{equation}
Such a distribution leads to:
\begin{eqnarray}
    F_{ci}(\Delta^2) &=&\frac{\sqrt{\pi}}{2}\sqrt{\frac{\Lambda^2_{ci}}{\Delta^2}} \,\text{erf}\left[ \sqrt{\frac{\Delta^2}{\Lambda^2_{ci}}}\right] \,,\\
    \theta_2^{{ci}}(\Delta^2)&=&2F_{ci}(\Delta^2)+\frac{\Lambda_{ci}^2}{\Delta^2}\left(e^{-\Delta^2/\Lambda_{ci}^2 }-1\right)\,,
\end{eqnarray}
with $\text{erf}[y]$ being the  Gaussian error function\,\cite{garfken67:math}. As can be noted, both form factors exhibit an asymptotic $1/\sqrt{\Delta^2}$ falloff. This reaffirms the fact that the form factors typically produced by the CI,\,\cite{Roberts:2010rn,Xing:2023eed,Zhang:2020ecj}, are harder than the ones predicted by QCD. The charge and the mass radii are fully determined by the mass scale $\Lambda_{ci}$, \emph{i.e.}:
\begin{equation}
    (r_E^{{ci}})^2=\frac{2}{\Lambda_{ci}^2}\,,\,(r_M^{{ci}})^2=\frac{1}{\Lambda_{ci}^2}\,\Rightarrow \mathcal{R}^{{ci}}_0=\frac{1}{2}\,.
\end{equation}
We would like to point out that the rightmost result is not a coincidence and, in fact, it can also be traced back to Eq.\,\eqref{eq:ratiofrac}.

Turning our attention to the IPS-GPD, a rather simple expression can readily be deduced:
\begin{equation}
    \label{eq:IPSci}
    \mathcal{I}_{ci}(x,|b_\perp|)=\frac{\Lambda_{ci}^2}{2(1-x)^2}|b_\perp|\,\text{exp}\left[-\frac{\Lambda_{ci}^2}{4(1-x)^2}|b_\perp|^2 \right]\,.
\end{equation}
This profile is maximal at $(x^{\text{max}},|b_\perp|^{\text{max}})\to(1,0)$. The corresponding charge and mass distributions are:
\begin{eqnarray}
    \rho_E^{{ci}}(|b_\perp|)&=&\frac{\Lambda_{ci}}{4|b_\perp| \sqrt{\pi}} \text{erfc}\left[\frac{1}{2} \Lambda_{ci} |b_\perp| \right]\,,\\
     \rho_M^{{ci}}(x,|b_\perp|)&=&\frac{1}{2}\rho_E^{{ci}}(|b_\perp|) + \frac{\Lambda_{ci}^2}{4\pi}\text{Ei}\left[-\frac{1}{4}\Lambda_{ci}^2|b_\perp|^2 \right]\,,\nonumber
\end{eqnarray}
where $\text{erfc}[z]=1-\text{erf}[z]$ and $\text{Ei}[z]$ is the exponential integral function\,\cite{garfken67:math}. Lastly, it is also important to note that both distributions exhibit a  $\sim 1/|b_\perp|$ divergence as $|b_\perp|\to 0$, which arises naturally from the $1/\sqrt{\Delta^2}$ asymptotic behavior of the form factors.

\subsection{Infinitely heavy-systems}
\label{sec:Infinitely}
Consider now the following DF:
\begin{equation}
    \label{eq:DFinfinity}
    q_{{nr}}(x)=\delta(x-1/2)\,,
\end{equation}
characterized by the moments:
\begin{equation}
\label{eq:MomsNR}
    \langle x^n \rangle_{q_{{nr}}}=\int_0^1 dx \, x^n q_{{nr}}(x) = \frac{1}{2^n}\,.
\end{equation}
This simple scenario leads us to the valence-quark GPD:
\begin{equation}
    H_{{nr}}^q(x,\Delta^2)= \delta(x-1/2) \, \text{exp}\left[-\frac{\Delta^2}{\Lambda_{nr}^2}(1-x)^2 \right]\,.
\end{equation}
Intuitively, the $q_{{nr}}(x)$ DF indicates with complete certainty that the valence-quark and antiquark share exactly half of the total momentum. This limiting case can be linked to a mesonic system made up of infinitely massive valence constituents. Consequently, the mass scale characterizing the system, $\Lambda_{{nr}}$, is expected to approach an asymptotically large value. Therefore, we have $H_{{nr}}^q(x,\Delta^2)\to  \delta(x-1/2)$, and
\begin{equation}
    F_{{nr}}(\Delta^2)= \theta_2^{{nr}}(\Delta^2)\to 1\,\Rightarrow\,(r_{E,M}^{nr})^2\to 0\,;\,\mathcal{R}_0^{{nr}}=1\,.
\end{equation}
As a result, the IPS-GPD also acquires a trivial representation ($\tilde{b}=\frac{1}{2}\Lambda_{\textbf{P}} |b_\perp|$):
\begin{eqnarray}
    \nonumber
    \mathcal{I}_{nr}(x,|b_\perp|) &=& \hspace{-1mm} \lim_{\Lambda_{nr}^2\to\infty} \hspace{-3mm} \delta(x-1/2)\frac{\Lambda_{nr}\tilde{b}}{(1-x)^2}\text{exp}\left[-\frac{\tilde{b}^2}{(1-x)^2} \right]\\
    &\to& \delta(x-1/2)\delta(|b_\perp|)\,.\label{eq:IPSnr}
\end{eqnarray}
Intuitively, this indicates that the valence-quark lies at the center of transverse momentum, carrying $1/2$ of the total momentum, \emph{i.e.}, $(x^{\text{max}},|b_\perp|^{\text{max}})\to(1/2,0)$. The associated charge and mass distributions are then:
\begin{equation}
\label{eq:rhoInfinity}
\rho_{E,M}^{{nr}}(|b_\perp|)=\lim_{\Lambda_{nr}^2\to\infty}\frac{1}{\pi} \Lambda_{nr}^2 e^{-\Lambda_{nr}^2|b_\perp|^2} \to \frac{1}{2\pi |b_\perp|}\delta(|b_\perp|)\,.
\end{equation}
Evidently, the profiles examined in this section suggest a compression of the infinitely massive system, ultimately approaching the point-like limit. In  all realistic systems, mesons exhibit spatial contraction as the mass of their constituent particles increases\,\cite{Raya:2024ejx,Xu:2024vkn,Arifi:2024mff,Li:2022izo,Chen:2018rwz,Li:2015zda}.

\subsection{Scale-free profile}
\label{sec:sf}
Consider now the usually-dubbed scale-free DF 
\begin{equation}
    \label{eq:PDFsf}
    q_{sf}(x)=30x^2(1-x)^2\,,
\end{equation}
which results from applying Eq.\,\eqref{eq:facPDFPDA} to the asymptotic limit of the pseudoscalar DA; namely\,\cite{Lepage:1980fj},
\begin{equation}
\varphi_{asy}(x):=6x(1-x)  \;;  
\end{equation}
and whose Mellin moments are:
\begin{equation}
    \langle x^n \rangle_{q_{sf}}=\int_0^1 dx\,x^nq_{sf}(x)=\frac{60}{(3+n)(4+n)(5+n)}\,.
\end{equation}
The valence-quark GPD then reads:
\begin{equation}
    \label{eq:GPDsf}
    H_{sf}^q(x,\Delta^2)=30x^2(1-x)^2 \,\text{exp} \left[-\frac{\Delta^2}{\Lambda_{\textbf{P}}^2}(1-x)^2 \right]\,.
\end{equation}
Albeit a bit larger than in previous cases, the corresponding EFFs and GFFs also undertake algebraic forms:
\begin{eqnarray}
\label{eq:EFFsf}
    F_{sf}(\Delta^2)&=&\frac{15\Lambda_{sf}^2}{\Delta^4}\left\{2\Lambda_{sf}^2(e^{-\Delta^2/\Lambda_{sf}^2}-4)\right. \\
 &+&   \left. 
    \sqrt{\pi} \sqrt{\frac{\Lambda_{sf}^2}{\Delta^2}}(3\Lambda_{sf}^2+4 \Delta^2)\text{erf}\left[ \sqrt{\frac{\Delta^2}{\Lambda_{sf}^2}}\right] \right\}\,,\nonumber\\
    \label{eq:GFFsf}
    \theta_2^{sf}(\Delta^2)&=&\frac{15\Lambda_{sf}^2}{\Delta^4}\left\{ 4\frac{\Lambda_{sf}^2}{\Delta^2}(4\Lambda_{sf}^2+\Delta^2)e^{-\Delta^2/\Lambda_{sf}^2}\right.\\
    &-&\left. 8\frac{\Lambda_{sf}^2}{\Delta^2}(2\Lambda^2+3\Delta^2)\right. \nonumber\\
    &+& \left. 2\sqrt{\pi} \sqrt{\frac{\Lambda_{sf}^2}{\Delta^2}}(9\Lambda_{sf}^2+2\Delta^2)\text{erf}\left[\sqrt{\frac{\Delta^2}{\Lambda_{sf}^2}}\right]\right\}\,.\nonumber
\end{eqnarray}
For their part, the charge and mass radii are expressed as:
\begin{equation}
\label{eq:R0asym}
(r_E^{sf})^2=\frac{12}{7\, \Lambda_{sf}^2}\,,\,(r_M^{sf})^2=\frac{9}{7\, \Lambda_{sf}^2}\,\Rightarrow \, \mathcal{R}_0^{sf}=\frac{3}{4}\,.
\end{equation}
The final outcome indicates that $\mathcal{R}_0^{sf}$ falls precisely in the middle point of $\mathcal{R}_0^{ci}=1/2$ and $\mathcal{R}_0^{nr}=1$. This is understandable as the profile of $q_{sf}(x)$ is strictly governed by counting rules,  without any broadening or dilation effects from mass generation mechanisms. 

Note that the EFF and GFF in Eq.\,\eqref{eq:EFFsf} and \eqref{eq:GFFsf}, exhibit a $1/(\Delta^2)^2$ suppression, exceeding the asymptotic expectations by an additional power of $\Delta^2$. This  is typical of simple GPD models, \emph{e.g.} those found in Refs.\,\cite{Mezrag:2016hnp,Chouika:2017rzs,Raya:2021zrz,Albino:2022gzs}. For practical purposes, this does not represent a setback. 

The IPS-GPD follows immediately,
\begin{eqnarray}
    \nonumber
    \mathcal{I}_{sf}(x,|b_\perp|) &=& 15x^2\Lambda_{sf}^2|b_\perp|\,\text{exp}\left[-\frac{\Lambda_{sf}^2}{4(1-x)^2}|b_\perp|^2 \right]\,,\label{eq:IPSsf}
\end{eqnarray}
and features its global maximum at 
\begin{equation}
(x^{\text{max}},|b_\perp^{\text{max}}|)=(2/3, \sqrt{2}/(3\Lambda_{sf}))\,.
\end{equation}
The deviation with respect to the previous cases reflects a more realistic dynamics.
The charge and mass distributions read:
\begin{eqnarray}
\label{eq:chargeDistsf}
    \rho_E^{sf}(|b_\perp|) &=& \frac{5\Lambda_{sf}^2}{2\pi} \left\{ e^{-\tilde{b}^2}(1-2\tilde{b}^2)\right.\\   
    &+& \left. \tilde{b}\left[(2\tilde{b}^2-3)\sqrt{\pi} \,\text{erf}[\tilde{b}]-3 \tilde{b} \,\text{Ei}[-\tilde{b}^2] \right] \right\}\,,\nonumber\\
\label{eq:massDistsf}
    \rho_M^{sf}(|b_\perp|) &=& \frac{3}{2}\frac{5\Lambda_{sf}^2}{2\pi} \left\{ e^{-\tilde{b}^2}(1-7\tilde{b}^2)\right.\\
    &+& \left. \tilde{b}\left[(8\tilde{b}^2-4)\sqrt{\pi}\, \text{erfc}[\tilde{b}]+(\tilde{b}^3-6\tilde{b})\,\text{Ei}[-\tilde{b}^2] \right]\right\},\nonumber
\end{eqnarray}
which turn to be finite as $|b_\perp|\to 0$:
\begin{eqnarray}\label{eq:sfRhoE0}
    \rho_E^{sf}(|b_\perp|) &\overset{|b_\perp|\to 0}{\approx}& \frac{5\Lambda_{sf}^2}{2\pi} \left[1-\frac{3}{2} \sqrt{\pi} |b_\perp| + \mathcal{O}(|b_\perp|^2)\right]\,,\hspace{0.7cm}\\
    \rho_M^{sf}(|b_\perp|) &\overset{|b_\perp|\to 0}{\approx}& \frac{3}{2}\rho_E^{sf}(|b_\perp|\to0)\,. \label{eq:sfRhoM0}
\end{eqnarray}

Briefly recapitulating,  $q_{{ci}}(x)$ and $q_{{nr}}(x)$ are the broadest and narrowest possible DFs, respectively, thereby understood as limiting cases. For its part, $q_{{sf}}(x)$ lies right in between, marking a  boundary between strong and weak mass generation dominance. In the case of pion-like systems, the expressions above enable us to set boundaries for the Mellin moments \,\cite{Lu:2023yna,Cui:2022bxn}:
\begin{equation}
\label{eq:bounds1}
    \frac{1}{2^n} \leq \langle x^n\rangle_{q_{\textbf{P}}} \leq \frac{1}{n+1}\,;
\end{equation}
and for the mass-to-charge ratio\,\cite{Xu:2023bwv}:
\begin{equation}
    \frac{1}{2} \leq \mathcal{R}_0^\textbf{P} \leq 1\,. 
\label{eq:bounds2}
\end{equation}
The upper limit of $\mathcal{R}_0^\textbf{P}$ holds mathematically for any positive and symmetric DF,\,\cite{Raya:2021zrz}, whereas the lower limit necessitates the use of the factorized LFWF prescription,\,\cite{Xu:2023bwv}. The validity of Eq.\,\eqref{eq:bounds2} has been examined from various perspectives, with a recent discussion available in Refs.\,\cite{Wang:2024sqg,Wang:2024fjt}.

Leaving these special cases behind for the moment, we now proceed to discuss the construction of realistic GPDs for pseudoscalar mesons.

\begin{figure*}[t]
\centerline{%
\begin{tabular}{ccc}
\includegraphics[width=0.3\textwidth]{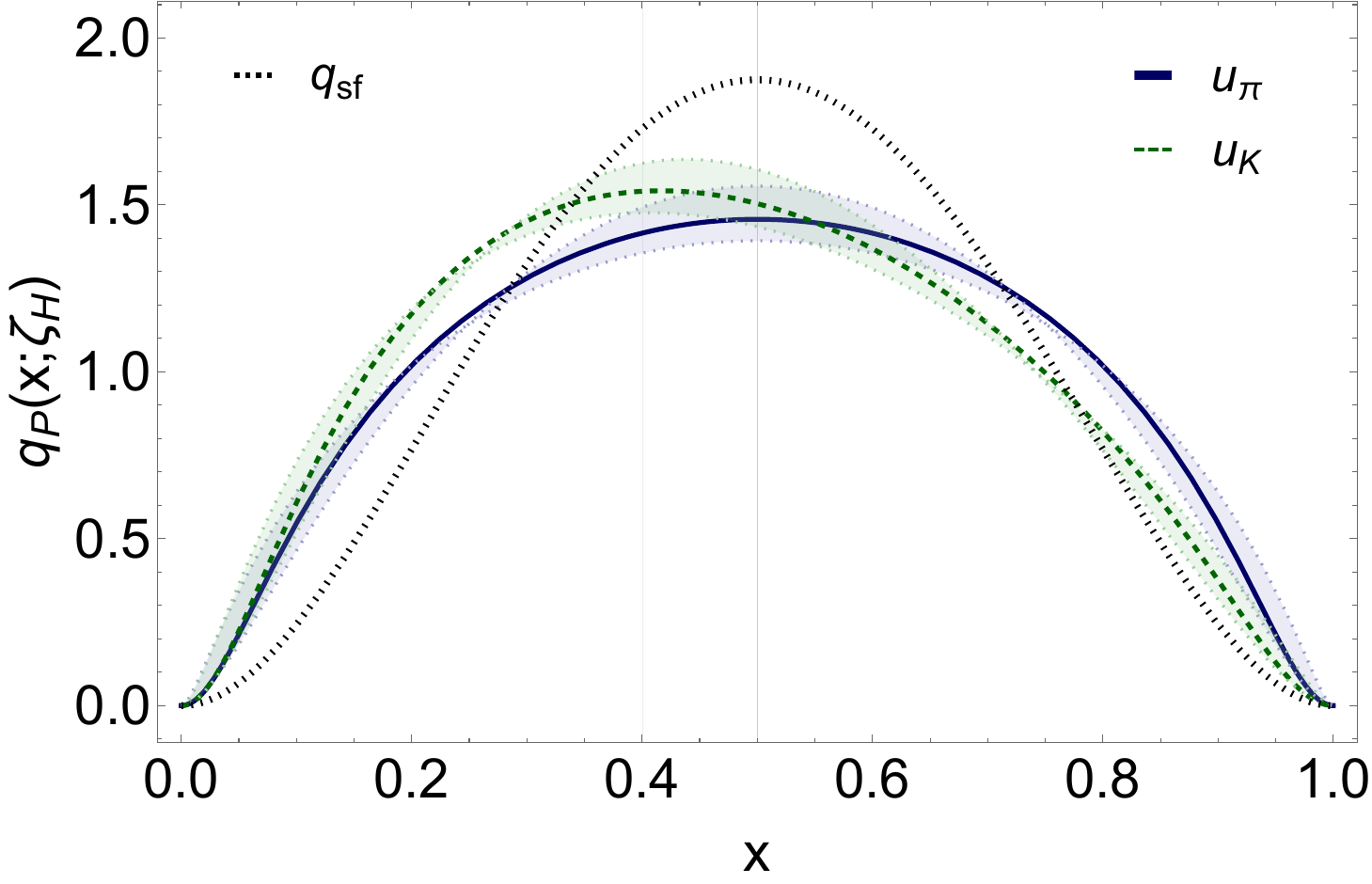} & \includegraphics[width=0.3\textwidth]{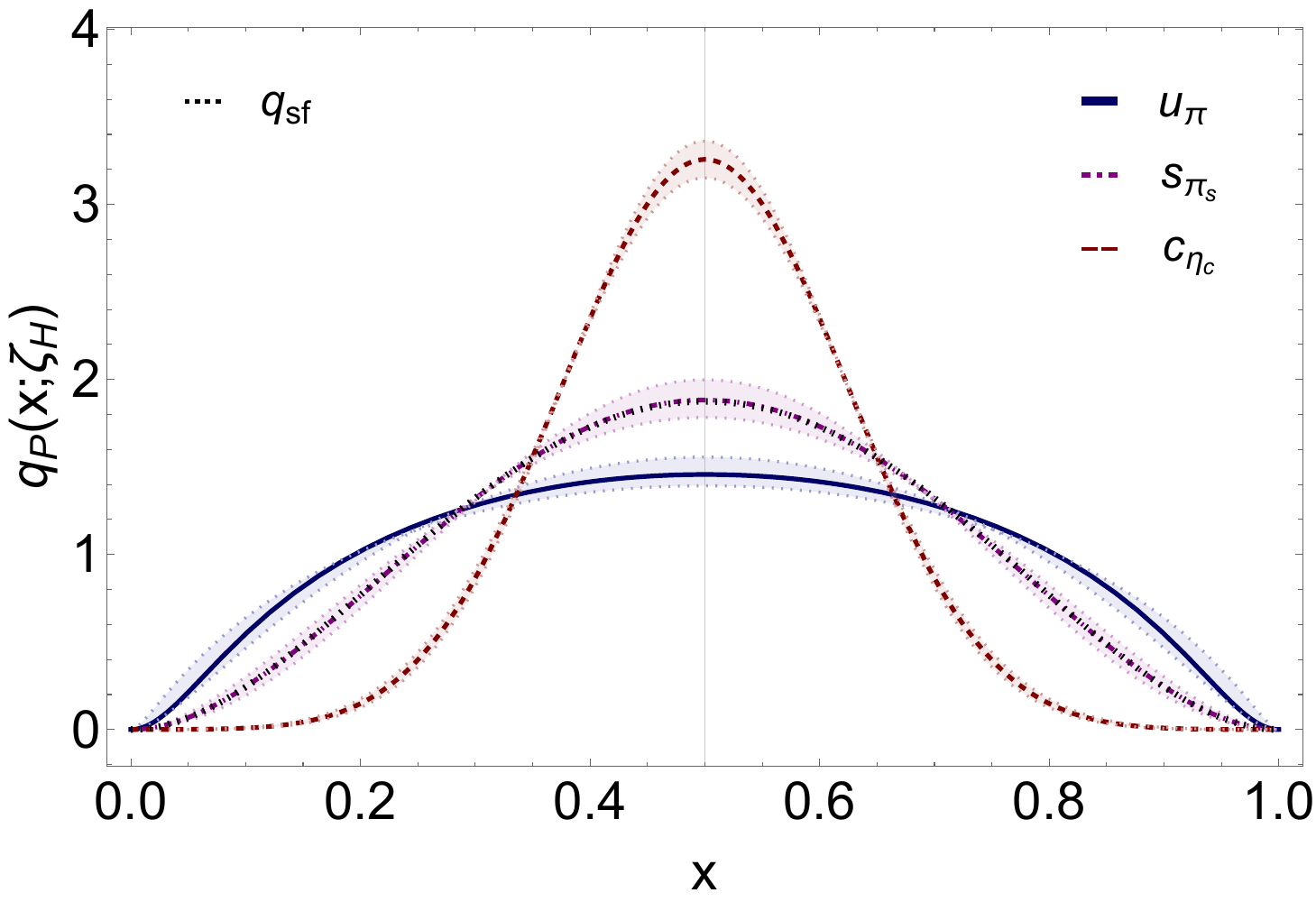} & \includegraphics[width=0.3\textwidth]{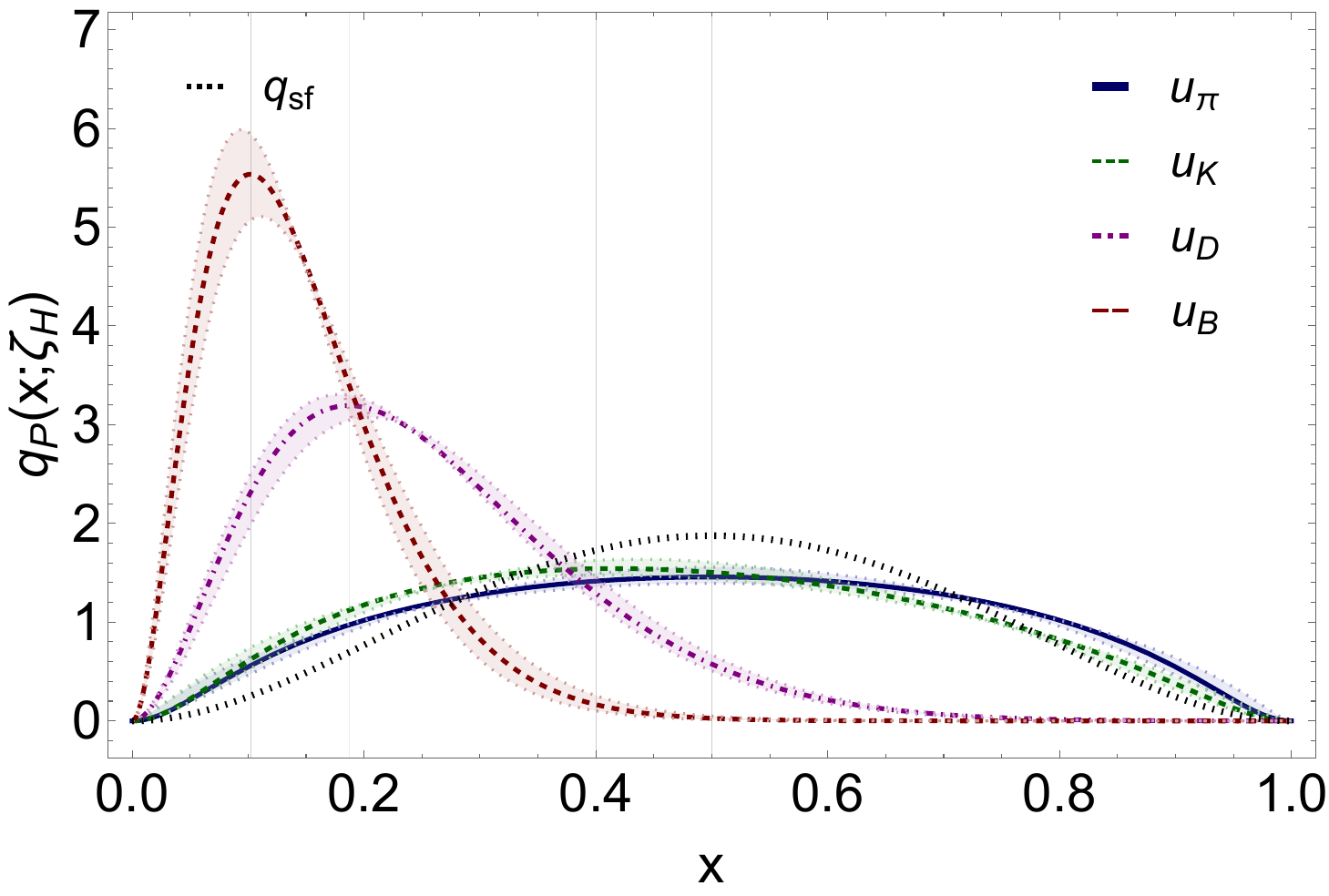}
\end{tabular}}
\caption{Valence-quark DFs of various ground-state pseudoscalar mesons, evaluated at $\zeta_H$. [left] Light mesons, \emph{i.e.} $\pi-K$. [center] Quarkonia systems. [right] Open-flavor states. The scale-free DF, Eq.\,\eqref{eq:PDFsf}, is shown for comparisson in every case. The vertical thin lines indicate the maximum of the given DF, and the error bands emanate from the associated DAs\,\cite{Raya:2021zrz}.}
\label{fig:DFs}     
\end{figure*}

\section{Pseudoscalar mesons GPDs}
\label{sec:numericalRes}

\subsection{Setting the stage}
Let us consider the valence-quark GPD in the ERS, Eq.\,\eqref{eq:GPDgen}, in combination with the profile function suggested in Eq.\,\eqref{eq:fxprofile}:
\begin{equation}
    \label{eq:LightPS}
    H_\textbf{P}^{\rm light}(x,\Delta^2)=q_\textbf{P}^{\rm light}(x) \,\text{exp} \left[-\frac{\Delta^2}{\Lambda_\textbf{P}^2}(1-x)^2 \right]\,.
\end{equation}
As outlined in the preceding sections, this form  can be linked to a LFWF in which the $x-k_\perp$ dependence is entirely factorized, as shown in Eq.\,\eqref{eq:PsiFac}. In this context, the GPD representation above is  applicable for mesons with masses $m_\textbf{P}\lesssim m_{\pi_s}\approx 0.69$ GeV\,\cite{Raya:2024ejx}. The  deviation from factorization is expected to be minimal in this case, see Eq.\,\eqref{eq:violation}. What remains to determine the GPD completely is an appropriate input for the corresponding DFs. We shall consider the $\pi-K$ DFs computed in\,\cite{Cui:2020tdf}, that can be well represented as follows\,\cite{Raya:2024ejx}:
\begin{equation}
\label{eq:PDFlog}
    q_\textbf{P}(x;\zeta_H) := n_\textbf{P}\ln \left[ 1 + \frac{x^2(1-x)^2}{\rho^2_\textbf{P}}\right](1+\gamma_\textbf{P} (1-2x))\,,
\end{equation}
with $n_\textbf{P}$ being a normalization constant ensuring baryon number conservation, and
\begin{equation}
\rho_\pi=0.069\,, \gamma_\pi=0\,,\quad 
\rho_K=0.087\,, \gamma_K=0.295\,.
\end{equation}
Such parametrization effectively incorporates the EHM-induced dilation, soft endpoint behavior, and skewing when applicable. Moreover, when these distributions are evolved to the phenomenologically relevant scales, a precise agreement with experimental as well as lattice QCD results is obtained\,\cite{Cui:2020tdf}.  Therefore, aside from the explicit error bands, it makes no difference whether the DF construction is instead solely guided by phenomenology\,\cite{Lu:2023yna,Cui:2021mom,Cui:2022bxn}.
\\
\\
Despite the noticeably concise form of Eq.\,\eqref{eq:PDFlog}, a  comprehensive utilization of the algebraic properties of the ERS approach becomes impaired in this case. Thus we  resort to a Gegenbauer polynomial expansion as follows:
\begin{equation}
    \label{eq:PDFlight}
    q_{\textbf{P}}^{\text{light}}(x;\zeta_H):=30x^2(1-x)^2\left(1+\sum_{j=1}^{N_G} a_j^\textbf{P} C_{j}^{5/2}(1-2x)\right)\,.
\end{equation}
Clearly $N_G=0$ recovers all the scale-free related results from Sec.\,\ref{sec:sf}. In contrast, a brief analysis shows that $N_G=4$ is sufficient to accurately represent the $\pi-K$ DFs from\,\cite{Cui:2020tdf}. The corresponding coefficients $a_j$ are listed in Table\,\ref{tab:paramsPDFlight}, and the associated distributions are depicted in Fig.\,\ref{fig:DFs}.

\begin{table}[t]
\centering
\caption{Interpolation parameters for the $u$ valence-quark DFs within the $\pi-K$ mesons. The distributions are expressed as in Eq.\,\eqref{eq:PDFlight} and are valid at $\zeta_H$.
\label{tab:paramsPDFlight}}
\begin{tabular}[t]{c||c|c|c|c}
\hline
$\,\,\textbf{P}\,\,\,\,\,\,$ & $\,\,a_1^\textbf{P}\,\,\,$ & $\,\,a_2^\textbf{P}\,\,\,$ & $\,\,a_3^\textbf{P}\,\,\,$ & $\,\,a_4^\textbf{P}\,\,\,$\\
\hline
$\pi$ &  $0.0$ & $0.112(28)$ & $0.0$ & $0.018(8)$\\
$K$ &  $\,-0.077(4)\,$ & $\,0.097(28)\,$ & $\,-0.010(3)\,$ & $\,0.013(7)\,$ \\
\hline
\end{tabular}
\end{table}%

Regarding systems with heavy quarks, Eq.\,\eqref{eq:violation} anticipates the inadequacy of employing a separable LFWF,  indicating that the extent of factorization violation is linked to the degree of chiral and flavor symmetry breaking. As a result, in contrast with Eq.\,\eqref{eq:facPDFPDA}, the connection between the DA and the DF is less apparent. Acknowledging these points and leveraging the available collection of DAs\,\cite{Binosi:2018rht,Ding:2015rkn}, we propose the following relationship:
\begin{equation}
    \label{eq:PDFheavyAns}
    q_{\textbf{P}}^{\rm heavy}(x;\zeta_H)=n_{\varphi}\frac{[\varphi_\textbf{P}^{\rm heavy}(x;\zeta_H)]^2}{(1+x \bar{\delta}_{qh})-x(1-x)}\,,
\end{equation}
where $\bar{\delta}_{qh}=1$ if $m_q\neq m_h$ and zero otherwise. The structure of the denominator in Eq.\,\eqref{eq:PDFheavyAns} is similar to that found in Refs.\,\cite{Albino:2022gzs,Almeida-Zamora:2023bqb,Serna:2024vpn}, but does not introduce additional mass scales and is pole-free on the real axis. Furthermore, given the defining relation in Eq.\,\eqref{eq:PDAmodel}, it is straightforward to obtain an expression for the profile function\,:
\begin{equation}
\label{eq:fxheavy}
    \hat{\phi}_\textbf{P}^{\rm heavy}(x;\zeta_H):=\frac{(1-x)^2}{\Lambda_\textbf{P}^2}\frac{1}{(1+x \bar{\delta}_{qh})-x(1-x)}\,.
\end{equation}
Having determined the latter, the ERS presented in Section\,\ref{sec:ERS} is now complete for heavy systems, leading to the following GPD:
\begin{equation}
\label{eq:HeavyPSG}
    H_\textbf{P}^{\rm heavy}(x,\Delta^2):=q_{\textbf{P}}^{\rm heavy}(x) \,\text{exp}\left[-\Delta^2 \hat{\phi}_\textbf{P}^{\rm heavy}(x;\zeta_H) \right]\,.
\end{equation}
Once again, $\zeta_H$ is assumed as the intrinsic scale of the model. The full set of parameters for the DA is compiled from Refs.\,\cite{Binosi:2018rht,Ding:2015rkn} and summarized in\,\cite{Raya:2024ejx}. DF-related Mellin moments are listed in Table\,\ref{tab:momentsPDF} and the associated DFs are depicted in Fig.\,\ref{fig:DFs}. The collection of reported results include the  hypothetical $\pi_s$ pseudoscalar. Notably it exhibits a profile which is nearly identical to $q_{sf}(x)$.  The Mellin moments associated with the $\eta-\eta'$ DFs are also largely comparable to this profile. Further comparisons might be found in, \emph{e.g.}, Refs.\,\cite{Serna:2024vpn,Serna:2020txe,Arifi:2024mff,Tang:2019gvn,Lan:2019img}

\begin{table}[t]
\centering
\caption{Mellin moments of the distribution functions for pseudoscalar mesons at $\zeta_H$. The DFs correspond to the lightest valence-quark within the meson. The error bands are derived from the errors associated with the DAs (see Ref.\,\cite{Raya:2024ejx}). In the $\eta-\eta'$ system, $l=u/d$ represents the $u/d$-quark component, while $s$ denotes the strange quark.\label{tab:momentsPDF}}
\begin{tabular}[t]{c|c|c|c|c}
\hline
$\textbf{P}$ & $\langle x^1 \rangle$ & $\langle x^2 \rangle$ & $\langle x^3 \rangle$ & $\langle x^4 \rangle$ \\
\hline
$\pi$ &  $0.500$ & $0.300(4)$ & $0.200(6)$ & $0.143(7)$ \\
$K$ &  $\,\,0.472(2)\,\,$ & $\,\,0.269(3)\,\,$ & $\,\,0.172(4)\,\,$ & $\,\,0.119(4)\,\,$ \\
\hline
$\eta_{l}$ &  $0.500$ & $0.291(3)$ & $0.187(4)$ & $0.128(5)$ \\
$\eta_s$ &  $0.500$ & $0.289(3)$ & $0.184(4)$ & $0.125(5)$ \\
$\pi_s$ &  $0.500$ & $0.286(3)$ & $0.178(4)$ & $0.119(5)$ \\
$\eta'_{l}$ &  $0.500$ & $0.283(3)$ & $0.174(3)$ & $0.114(4)$ \\
$\eta'_s$ &  $0.500$ & $0.282(3)$ & $0.173(3)$ & $0.112(4)$ \\
\hline
$D$ &  $0.257(11)$ & $0.084(6)$ & $0.032(3)$ & $0.014(2)$ \\
$D_s$ &  $0.274(9)$ & $0.094(5)$ & $0.038(3)$ & $0.017(2)$ \\
$\pi_c/\eta_c$ &  $0.500$ & $0.265(1)$ & $0.147(1)$ & $0.085(2)$ \\
\hline
$B$ &  $0.147(11)$ & $0.029(4)$ & $0.007(1)$ & $0.002(1)$ \\
$B_s$ &  $0.152(9)$ & $0.030(4)$ & $0.007(2)$ & $0.002(1)$ \\
$B_c$ &  $0.2705(4)$ & $0.083(3)$ & $0.0279(3)$ & $0.010(2)$ \\
$\pi_b/\eta_b$ &  $0.500$ & $0.258(1)$ & $0.137(1)$ & $0.075(2)$ \\
\hline
\end{tabular}
\end{table}%

\begin{figure*}[t]
\centerline{%
\begin{tabular}{cc}
\includegraphics[width=0.35\textwidth]{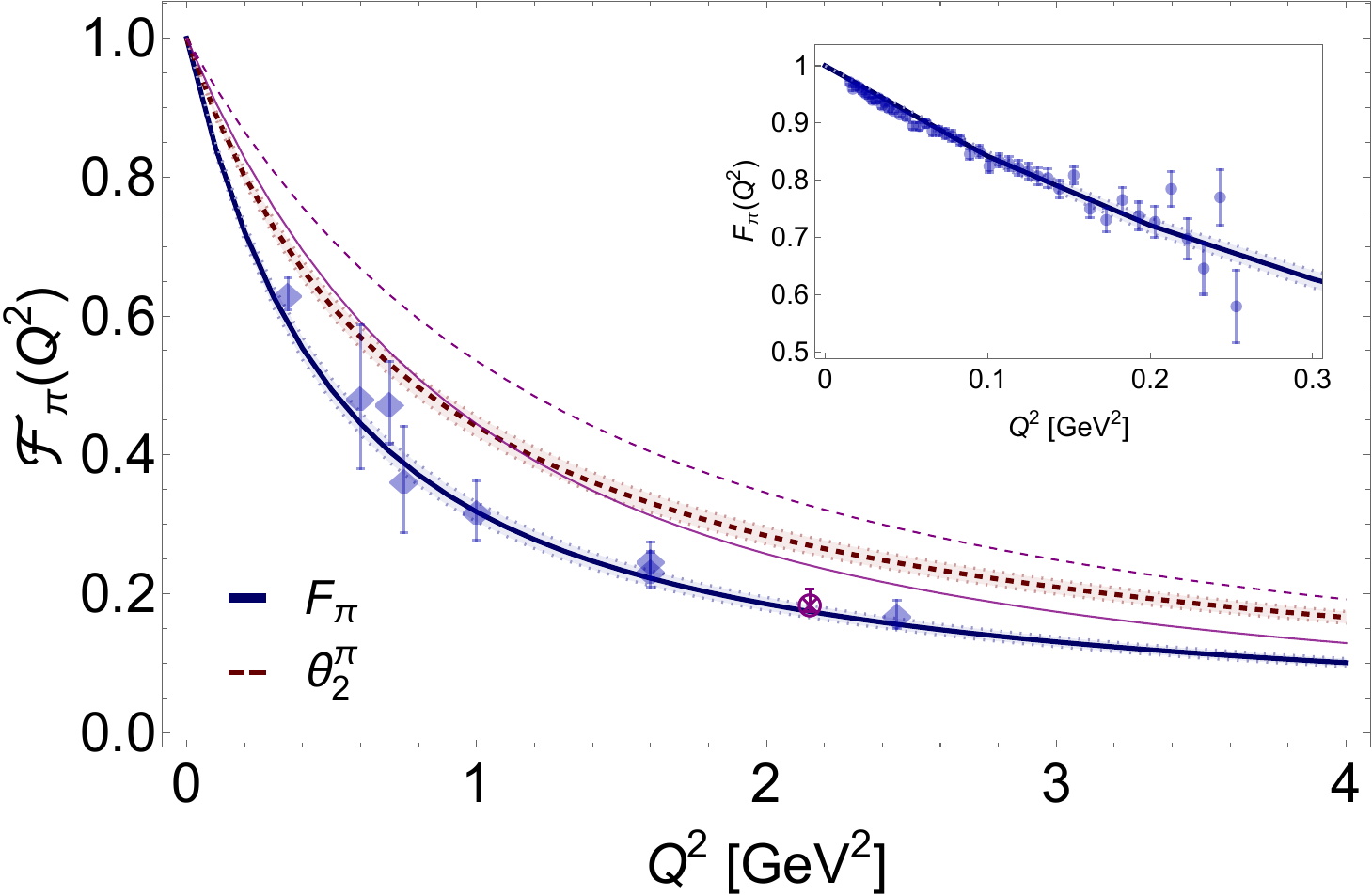}\hspace{0.3cm}\textbf{A} & \,\includegraphics[width=0.35\textwidth]{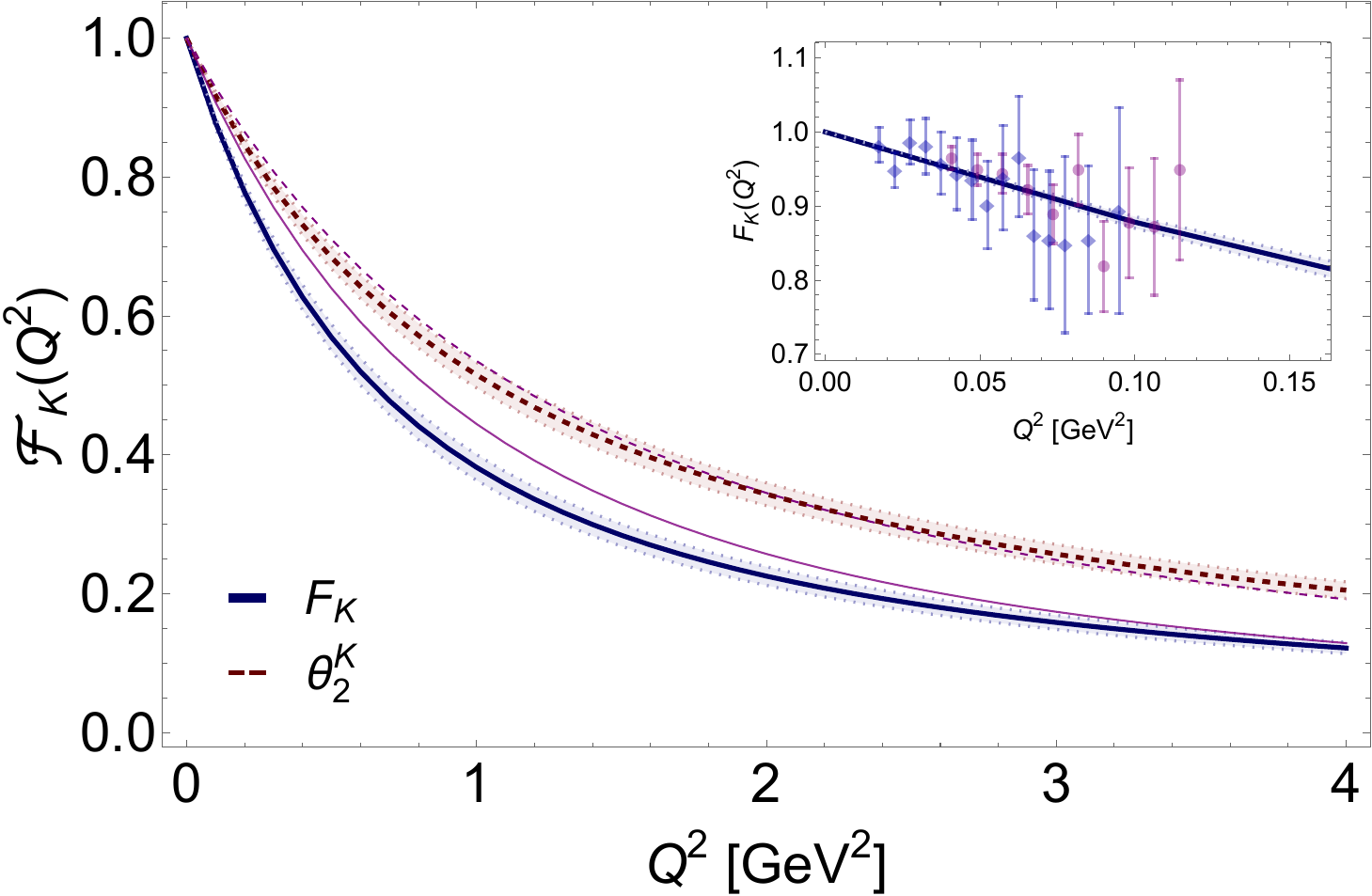}\hspace{0.3cm}\textbf{B} \\
\includegraphics[width=0.35\textwidth]{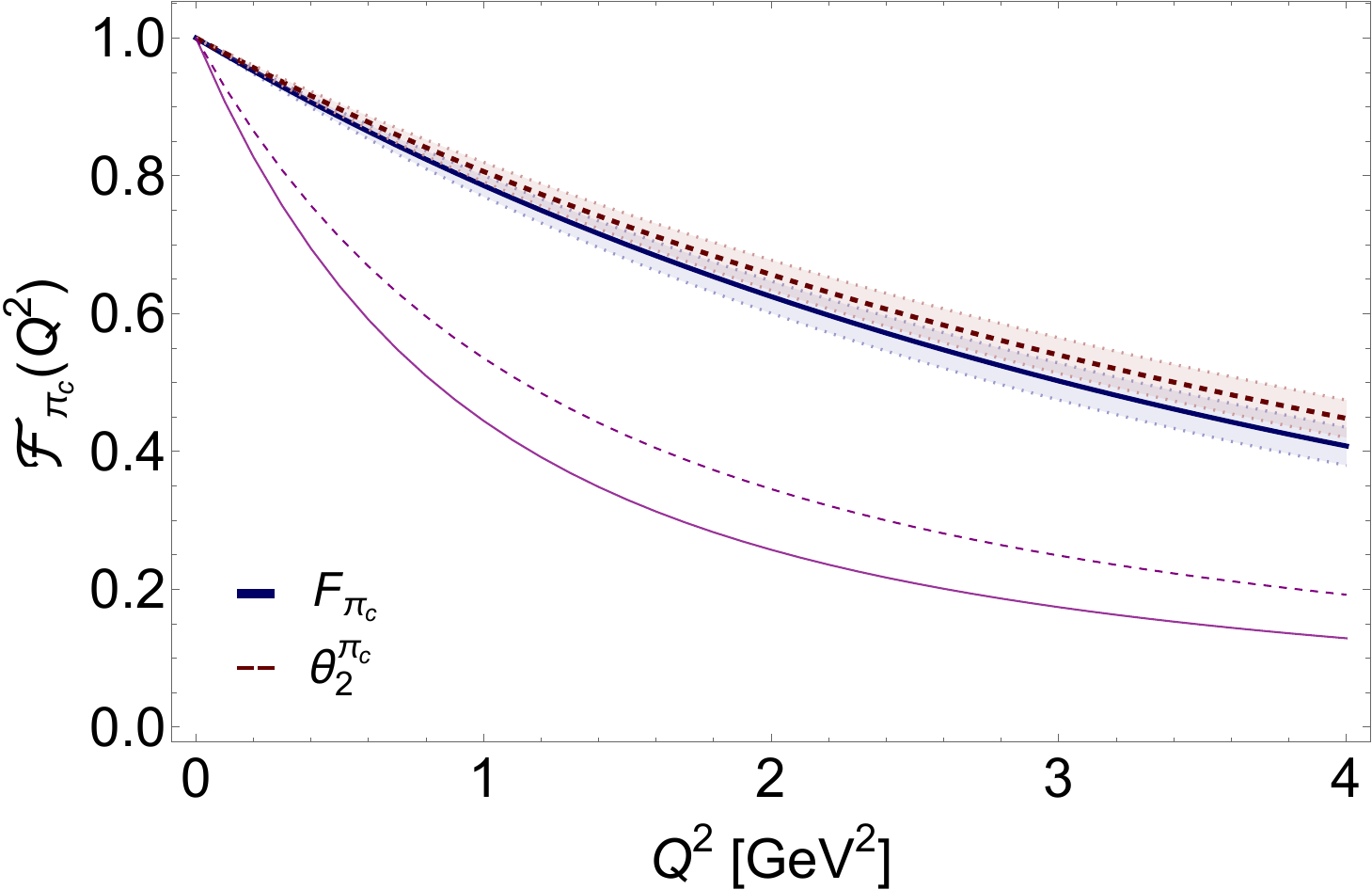}\hspace{0.3cm}\textbf{C} & \,\includegraphics[width=0.35\textwidth]{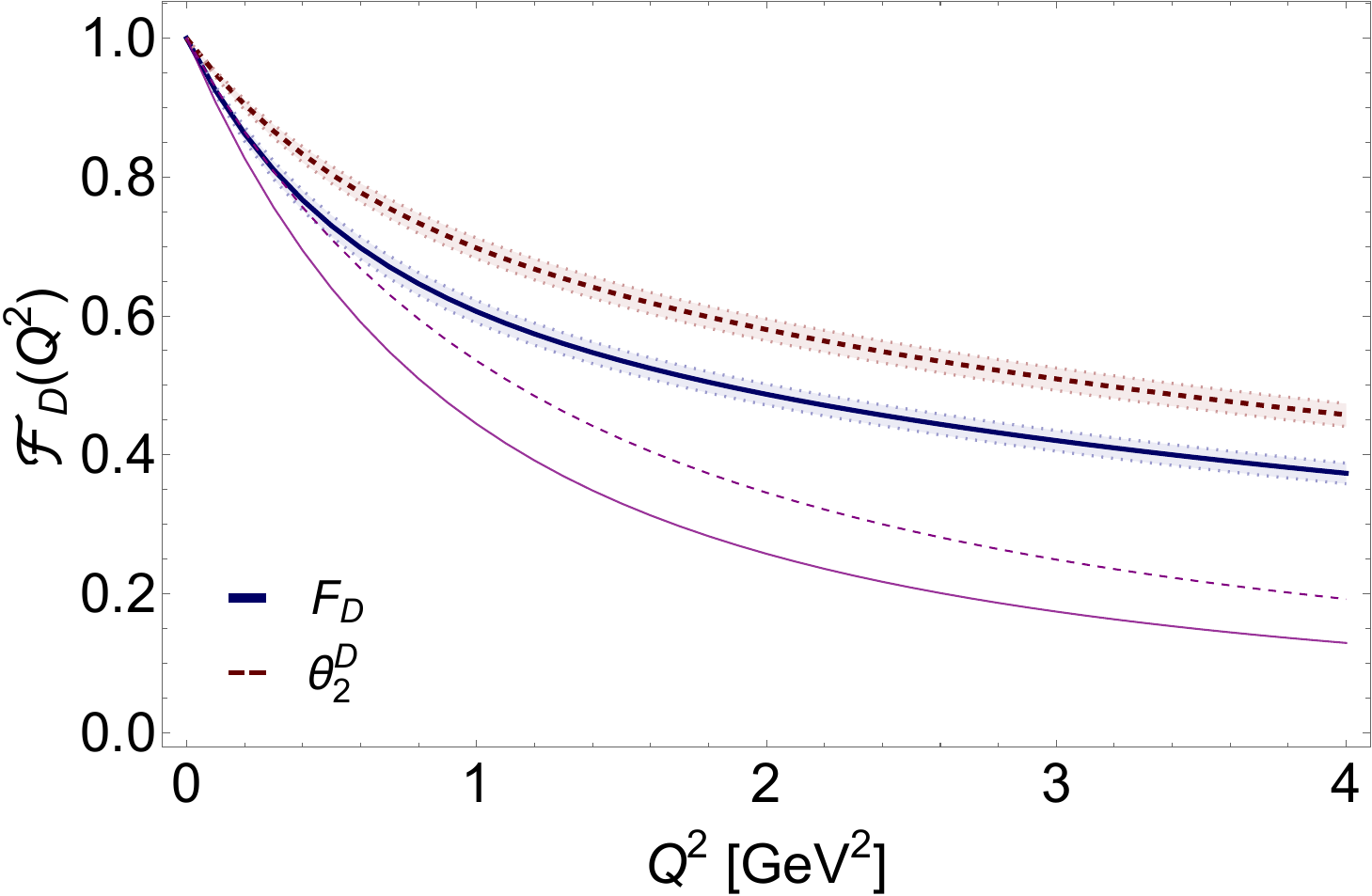}\hspace{0.3cm}\textbf{D}
\end{tabular}}
\caption{Electromagnetic and gravitational form factors (solid and dashed lines, respectively). [\textbf{A}] $\pi^+$. [\textbf{B}] $K^+$. [\textbf{C}] $\pi_c$. [\textbf{D}] $D^+$ meson. In each case, results for the $\pi_s$ pseudoscalar are included as thin purple lines for comparisson. Pion EFF experimental data from Refs.\,\cite{NA7:1986vav, Horn:2007ug, JeffersonLab:2008jve}; Kaon data from Refs.\,\cite{Dally:1980dj,Amendolia:1986ui}.}
\label{fig:FFs}     
\end{figure*}

\subsection{EFFs and GFFs}
It is reasonable to inquire whether the closed-form expressions for the electromagnetic and gravitational form factors can be derived, considering the more complex structure of $H_\textbf{P}^{\rm light}(x,\Delta^2)$ and $H_\textbf{P}^{\rm heavy}(x,\Delta^2)$, defined in Eq.\,\eqref{eq:LightPS} and Eq.\,\eqref{eq:HeavyPSG}, respectively. Only in the first case the answer is positive, though naturally, the complexity of the algebraic expressions depends on the value of $N_G$ — see Eq.\,\eqref{eq:PDFlight}.

Therefore, for the time being we shall focus on the light systems described by $H_\textbf{P}^{\rm light}(x,\Delta^2)$, starting by discussing the behavior of the form factors when approaching the $\Delta^2\to0,\,\infty$ limits. Considering first the charge-to-mass ratio $\mathcal{R}_{0}^{\textbf{P}}$, we arrive at:
\begin{equation}
\label{eq:radiiLightR}
    \mathcal{R}_0^\textbf{P}= \frac{9-5a_2^{\textbf{P}}}{12-5a_1^{\textbf{P}}+5a_2^{\textbf{P}}}\,,
\end{equation}
where the individual radii are:
\begin{eqnarray}
    \label{eq:radiiLight}   (r_E^{\textbf{P}})^2=\frac{12-5a_1^{\textbf{P}}+5a_2^{\textbf{P}}}{7\, \Lambda_{\textbf{P}}^2}\,,\,(r_M^{\textbf{P}})^2=\frac{9-5a_2^{\textbf{P}}}{7\, \Lambda_{\textbf{P}}^2}\,.
\end{eqnarray}
Neither value depend on the coefficients $a_j$ ($j\,>\, 2$) and, in addition, the mass radius is expressed solely by the even coefficients. At the opposing end,
\begin{equation}
   \mathcal{R}_\infty^\textbf{P}:=\lim_{\Delta\to\infty} \frac{\theta_2^{\textbf{P}}(\Delta^2)}{F_{\textbf{P}}(\Delta^2)}=\frac{6+90a_2^{\textbf{P}}+420a_4^{\textbf{P}}}{3+5a_1^{\textbf{P}}+45a_2^{\textbf{P}}+35 a_3^{\textbf{P}}+210a_4^{\textbf{P}}}\,.\label{eq:Ratinf}
\end{equation}
For any isospin-symmetric system, due to the vanishing odd coefficients $a_j$, $R_\infty^{\textbf{P}}=2$. This same outcome would be obtained when considering either the flat top or scale-free DFs discussed previously.

Capitalizing on the $\pi-K$ DFs considered here, 
\begin{equation}
\label{eq:ratiosRealPiK}
    \mathcal{R}_0^{\pi^+}=0.67(2)\,,\,
   \mathcal{R}_0^{K^+}=0.66(2)\,.
\end{equation}
Clearly, $\mathcal{R}_0^{\pi^+} = \mathcal{R}_0^{K^+}$ within the associated uncertainties. As has been discussed, the observation that both quantities are less than unity indicates a more compact mass distribution compared to that of charge. These findings are consistent with earlier studies, such as those in Refs.\,\cite{Raya:2024ejx,Xu:2023izo,Xu:2023bwv}. Focusing on $\mathcal{R}_0^\pi$, our calculated value aligns closely with the data-driven analysis in Ref.\,\cite{Xu:2023bwv}, which reports $\mathcal{R}_0^\pi = 0.62(5)$. Compatibility with other frameworks is also found. For instance, the LFHQCD framework yields $\mathcal{R}_0^\pi = 0.605$\,\cite{Wang:2024sqg}, while the GPD approach in\,\cite{Raya:2021zrz} estimates $\mathcal{R}_0^\pi = 0.659$.

\begin{figure*}[t]
\centerline{%
\begin{tabular}{ccc}
\includegraphics[width=0.3\textwidth]{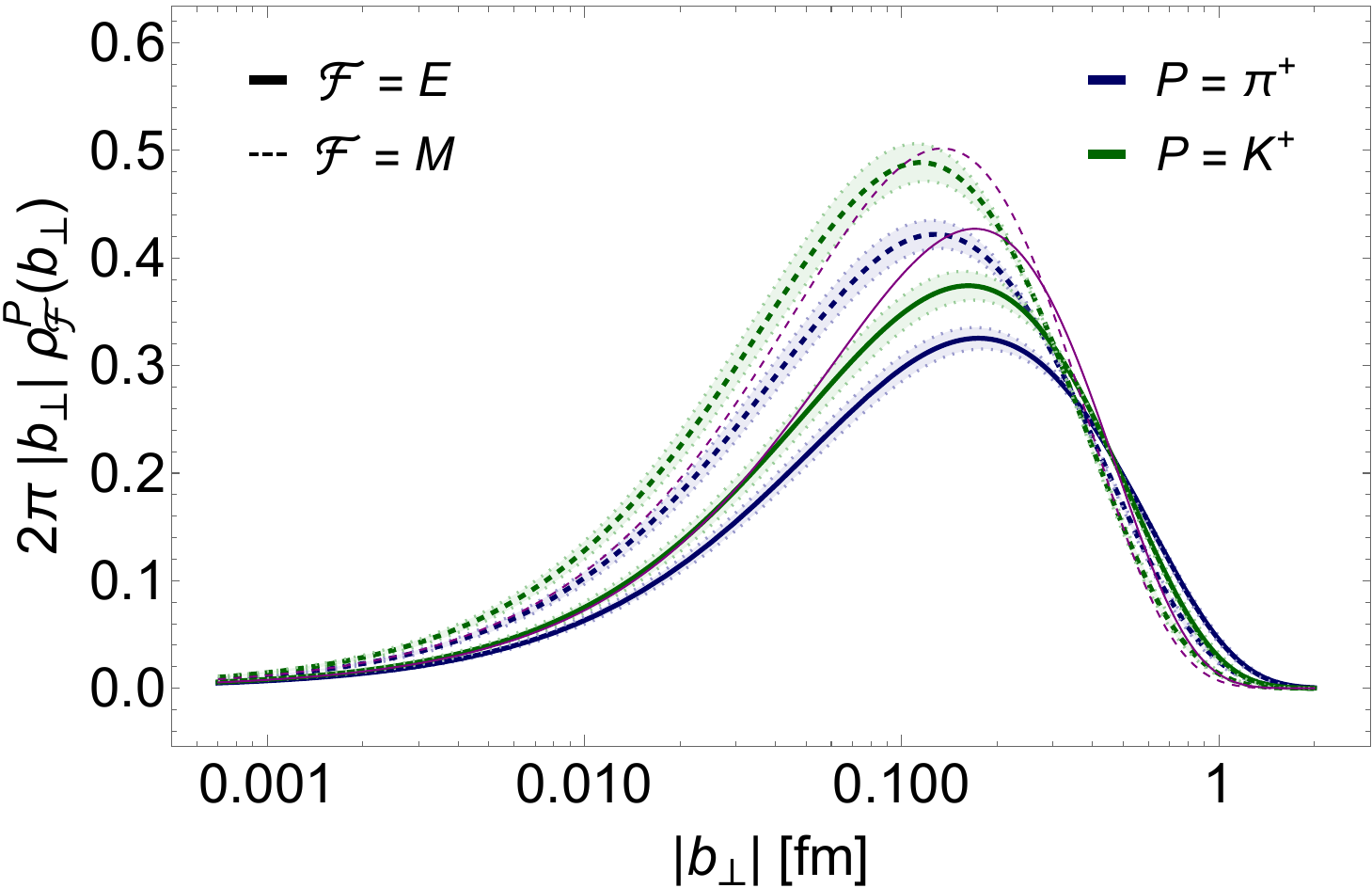} & \includegraphics[width=0.3\textwidth]{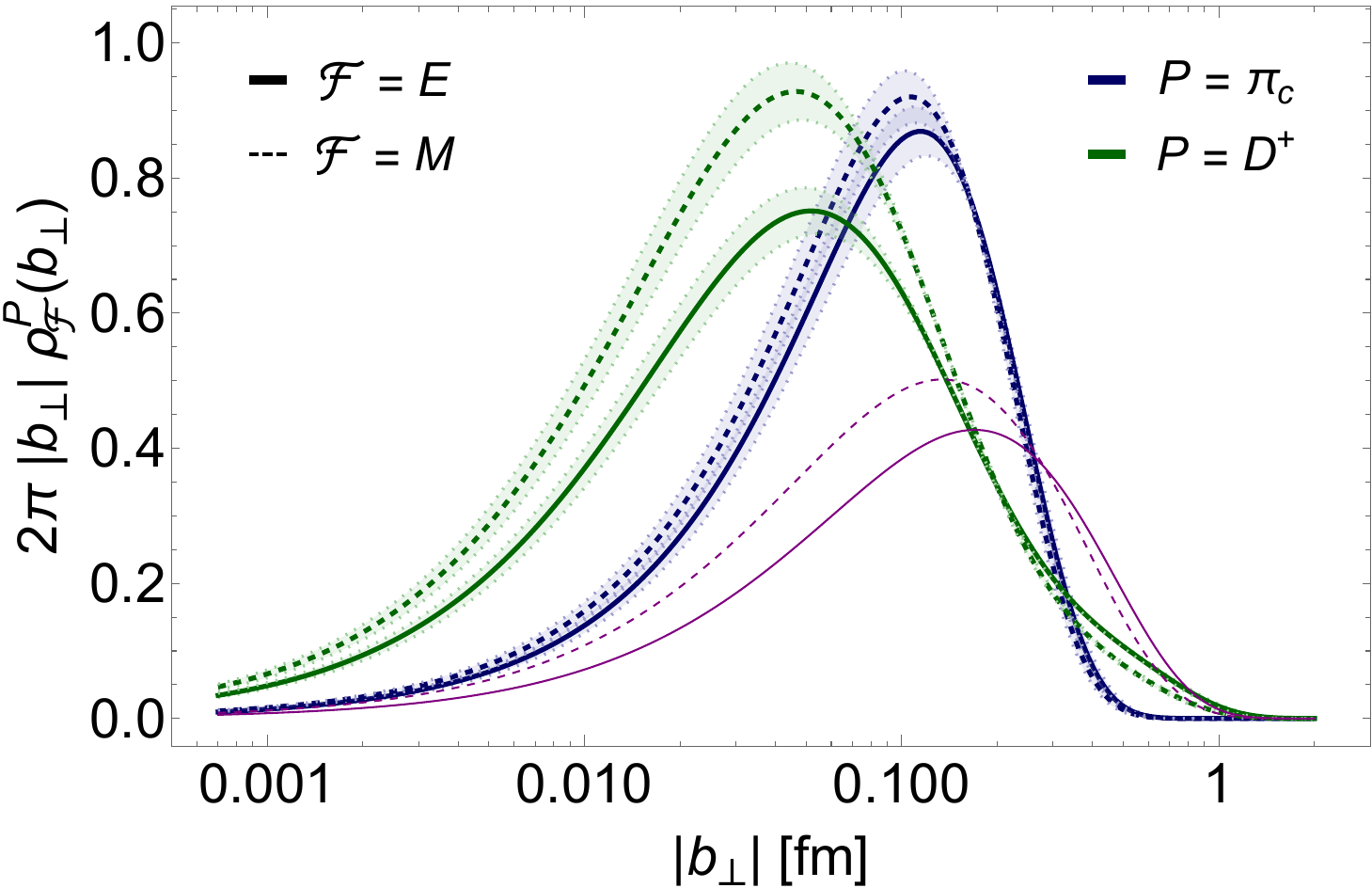} & \includegraphics[width=0.3\textwidth]{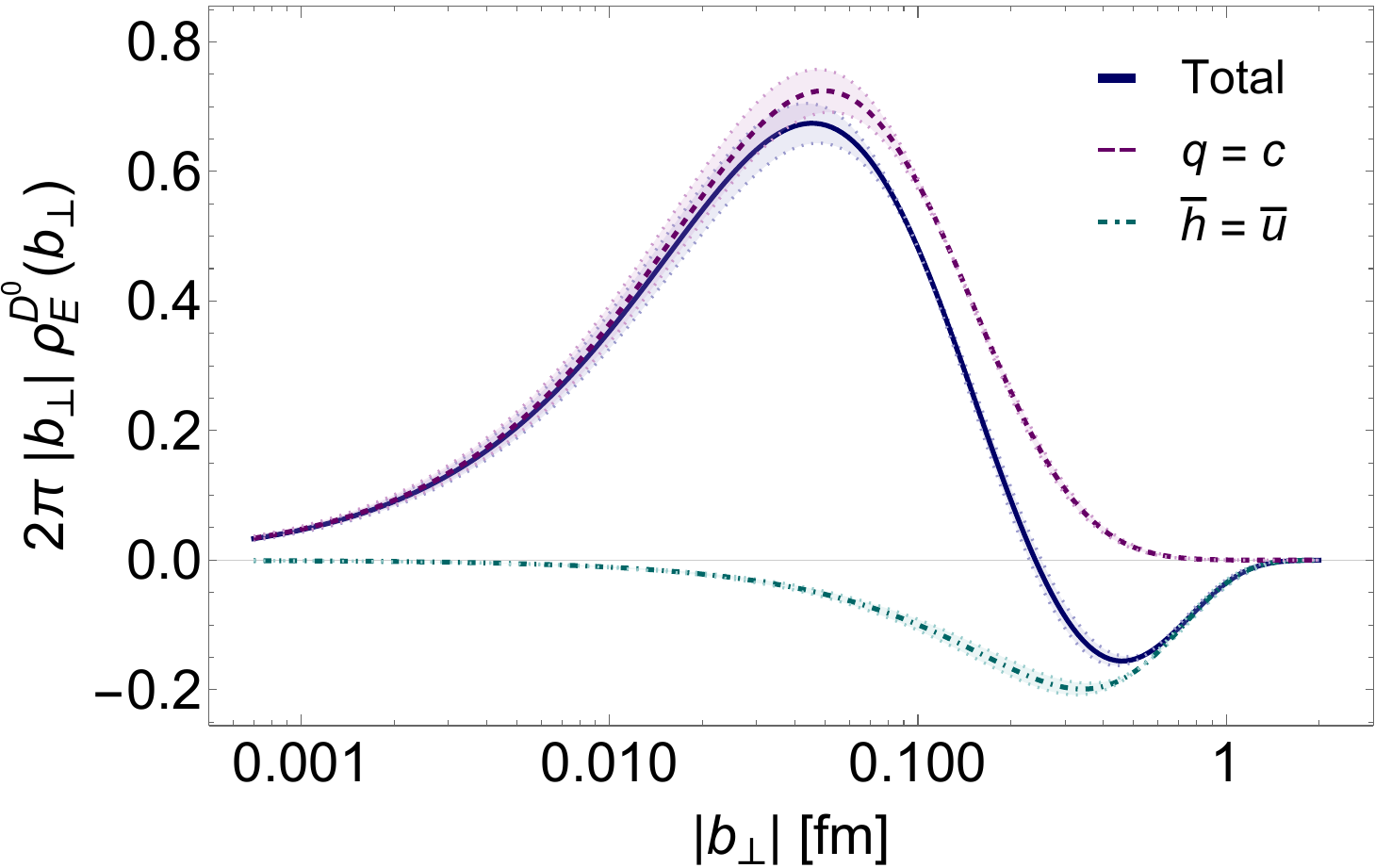}
\end{tabular}}
\caption{Charge and mass distributions as defined in Eq.\,\eqref{eq:chargeMass} (solid and dashed lines, respectively). [\textbf{left}] $\pi^+$ and $K^+$ distributions. [\textbf{center}] $\pi_s$ and $D^+$. [\textbf{right}] Flavor-separated $D^0=c\bar{u}$ meson: (dashed) $c$-quark and (dot-dashed) $\bar{u}$ antiquark. In each case, results for the $\pi_s$ pseudoscalar are included as thin purple lines for comparisson. }
\label{fig:Dists}     
\end{figure*}

Turning the attention to the large-$\Delta^2$ domain, one gets:
\begin{equation}
    \label{eq:ratiosPiK}
   \mathcal{R}_\infty^{\pi^+}=2.00(0)\,,\,\mathcal{R}_\infty^{K^+}=2.16(2)\,;
\end{equation}
being $ \mathcal{R}_\infty^{\pi^+},\mathcal{R}^{K^+}\gtrsim 2$, both results meet the expectations from pQCD\,\cite{Tong:2021ctu,Lepage:1980fj}.

To illustrate the pointwise behavior of the form factors, it is necessary to specify $\Lambda_{\pi,K}$. Driven by phenomenological analyses\,\cite{Cui:2022fyr,ParticleDataGroup:2024cfk}, we adopt $\Lambda_\pi=0.407(13)$ GeV and $\Lambda_K=0.456(16)$ GeV, in order to produce\,\footnote{Taking into account the variation with $\Lambda_\textbf{P}$, the errors in Eq.\,\eqref{eq:ratiosRealPiK} increase, thus producing $\mathcal{R}_0^{\pi^+,\,K^+}=0.67(7)\,,\,
   0.66(8)$. }:
\begin{equation}
    r_E^{\pi^+,K^+}=0.65(2),\,0.58(2)\, \text{fm}\,.
\end{equation}
The resulting $\pi-K$ electromagnetic and gravitational form factors are shown in the top-panels of Fig.\,\ref{fig:FFs}. For both mesons, the corresponding EFFs show precise agreement with the experimental data\,\cite{NA7:1986vav, Horn:2007ug, JeffersonLab:2008jve,Dally:1980dj, Amendolia:1986ui}. As far as GFFs are concerned, it is clear that their profiles are harder than those of the EFFs, thereby producing the following mass radii:
\begin{equation}
    r_M^{\pi^+,K^+}=0.53(2),\,0.47(2)\, \text{fm}\,.
\end{equation}
As can be  readily deciphered, whereas the mass-to-charge ratio for the pion and kaon is virtually identical, Eq.\,\eqref{eq:ratiosRealPiK}, the radius associated with $K^+$ is smaller in each case. In fact, when averaging $r_E^\textbf{P}$ and $r_M^\textbf{P}$, this contraction is of the order of $r^K/r^\pi=0.89(3)\approx f_\pi/f_K$. As a  comparative reference, calculation reported in\,\cite{Xu:2023izo} yields $r^K/r^\pi=0.85(6)$. Thus, despite its simplicity, the current scheme generates EFFs and GFFs that accurately reflect those obtained through the usage of more refined tools and reliable methods (see \emph{e.g.} \cite{Miramontes:2021exi,Chang:2013nia,Xu:2023izo}).

Further benchmarks are provided by the $\pi_s$ system. In this connection, guided by lattice QCD and CSM analyses\,\cite{Chen:2018rwz,Koponen:2017fvm}, we have selected $\Lambda_{\pi_s}=0.605(25)$ GeV, leading to (the errors considering the variation of $\Lambda_{\textbf{P}}$):
\begin{equation}
\label{eq:etasR0}
    r_E^{\pi_s}=0.48(2)\,\text{fm}\,,\,r_M^{\pi_s}=0.42(2)\,\text{fm}\,\Rightarrow \mathcal{R}^{\pi_s}_0=0.76(11)\,.
\end{equation}
Naturally, $\mathcal{R}_0^{\pi_s}\approx \mathcal{R}_0^{sf}=3/4$, given the similarity  between the DF of the $\pi_s$ and the scale-free profile, Eq.\,\eqref{eq:R0asym}. Moreover, as revealed by Fig.\,\ref{fig:FFs}, the form factors of the $\pi_s$ state are  harder than those of the pion and kaon; the notable exception is $\theta_2^{\pi_s}$ being comparable to the kaon's case. This naturally leads to mass distributions that exhibit some resemblance, as analyzed later on.

At the other end of the spectrum, for illustrative purposes, we present the $\pi_c$ and $D^+$ mesons in the bottom-panel of Fig.\,\ref{fig:FFs}. It is evident that the form factors of these systems are considerably harder than those of $\pi_s$ (consequently, from the $\pi-K$ too), highlighting the presence of a heavy quark in their valence structure.  Moreover, while the hierarchy $\theta_2^{\textbf{P}}(\Delta^2) > F_{\textbf{P}}(\Delta^2)$ is maintained, in the case of $\pi_c$, these quantities are nearly  the same. Note that the form factors for $\pi_c$ and $D^+$ are evaluated by determining the model’s only free parameter, $\Lambda_{\textbf{P}}$, based on the charge radii provided by lattice QCD and CSMs\,\cite{Xu:2024fun,Li:2017eic}. In this case, $\Lambda_{\pi_c}=1.183(49)$ GeV and $\Lambda_{D}=0.516(24)$ GeV produce:
\begin{eqnarray}\nonumber
    r_E^{\pi_c}=0.24(1)\,\text{fm}\,,\,r_M^{\pi_c}=0.23(1)\,\text{fm}\,\Rightarrow \mathcal{R}^{\pi_c}_0=0.92(11)\,;\hspace{0.65cm}&\\
    r_E^{D}=0.44(2)\,\text{fm}\,,\,r_M^{D}=0.37(2)\,\text{fm}\,\Rightarrow \mathcal{R}^{D}_0=0.71(12)\,.\hspace{0.7cm}&
\end{eqnarray}
The more noticeable proximity of $R_0^{\pi_c}$ to 1 is due to the fact that the two form factors are almost identical. It implies that the spatial distributions of charge and mass are similar. Therefore, these distributions have commensurate spatial extent. Exact equality would result in the non-relativistic limit. Compared to its value for the $\pi_s$ case, Eq.\,\eqref{eq:etasR0}, and despite the larger mass of the $D^+$, $\mathcal{R}_0^D$ turns out to be smaller for the latter. This is due to a non-trivial interplay between the weak and the strong mass generating mechanisms that takes place among the valence constituents of heavy-light systems. A more detailed analysis of the charge and mass distributions, along with the IPS-GPDs, will be covered in the following section.

\subsection{Spatial distributions}

Let us now plot and discuss the charge and mass distributions  given by Eq.\,\eqref{eq:chargeMass}. The corresponding results for the light systems, $\pi^+$ and $K^+$, are depicted in the left panel of Fig.\,\ref{fig:Dists}. Plainly the charge distribution extends slightly beyond the mass distribution, highlighting the larger charge radius relative to the mass one. The figure also includes the results for the pseudoscalar $\pi_s$, which follow the same trend.  Nonetheless, while its charge distribution is  narrower than that of the pion and the kaon, its mass distribution is on par with that of the latter.

The central panel of Fig.\,\ref{fig:Dists} illustrates what happens in the heavier systems $\pi_c$ and $D^+$. It is evident that the distributions for these mesons are sharply peaked at smaller values of $|b_\perp|$, as compared with the $\pi_s$ results employed for reference. In the case of the $\pi_c$, charge and mass distributions are rather similar, which is the trend for increasingly heavier systems. As we discussed in Section\,\ref{sec:Infinitely}, one would expect that as the mass of the bound state increases, the corresponding distributions would peak sharply at $|b_\perp|\to0$ (see Eq.\,\eqref{eq:rhoInfinity}). However, the case of the $D$ meson is  somewhat counterintuitive because, compared to those of $\pi_c$, its distributions are maximal at smaller values of $|b_\perp|$. This is attributed to the pronounced flavor asymmetry of this system. For a better understanding, let us consider the neutral state $D^0=c\bar{u}$, whose flavor-separated charge distribution is shown in the right panel of Fig.\,\ref{fig:Dists}.  Clearly, the heavier $c$-quark predominantly determines the peaked shape of the distributions in the small-$|b_\perp|$ domain. In this region, the $\bar{u}$-antiquark barely plays a role and only introduces some modulations at larger values of $|b_\perp|$. In the case of the charged state $D^+=c\bar{d}$, this interference is constructive and, therefore, it further enhances the peaked profile of the produced densities.

Another measure of the compression of the mass relative to charge profile is given by:
\begin{equation}
 \frac{\rho_M^{\pi^+}(|b_\perp|)}{\rho_E^{\pi^+}(|b_\perp|)}\overset{|b_\perp|\to 0}{\approx}1.66(3)\,,\; \;
 \frac{\rho_M^{K^+}(|b_\perp|)}{\rho_E^{K^+}(|b_\perp|)}\overset{|b_\perp|\to 0}{\approx}1.74(3)\,.
\end{equation}
\\
Our computed results exceed the partonic limit prediction inferred from Eqs.\,\eqref{eq:sfRhoE0} and \eqref{eq:sfRhoM0}, \emph{i.e.} $3/2$. This is nothing but a consequence of the EHM-induced dilation featured by the DFs. Concerning the heavy states, one finds:
\begin{equation}
 \frac{\rho_M^{\pi_c}(|b_\perp|)}{\rho_E^{\pi_c}(|b_\perp|)}\overset{|b_\perp|\to 0}{\approx}1.16(5)\,,\,
 \frac{\rho_M^{D^+}(|b_\perp|)}{\rho_E^{D^+}(|b_\perp|)}\overset{|b_\perp|\to 0}{\approx}1.36(5)\,,
\end{equation}
which lie closer to unity, \emph{i.e.}, the expectation for the infinitely-heavy system, Eq.\,\eqref{eq:rhoInfinity}. Unsurprisingly,  
\begin{equation}
 \frac{\rho_M^{\pi_s}(|b_\perp|)}{\rho_E^{\pi_s}(|b_\perp|)}\overset{|b_\perp|\to 0}{\approx}1.53(4)\,,
\end{equation}
right above the partonic limit case.
\begin{figure*}[t]
\centerline{%
\begin{tabular}{cc}
\includegraphics[width=0.35\textwidth]{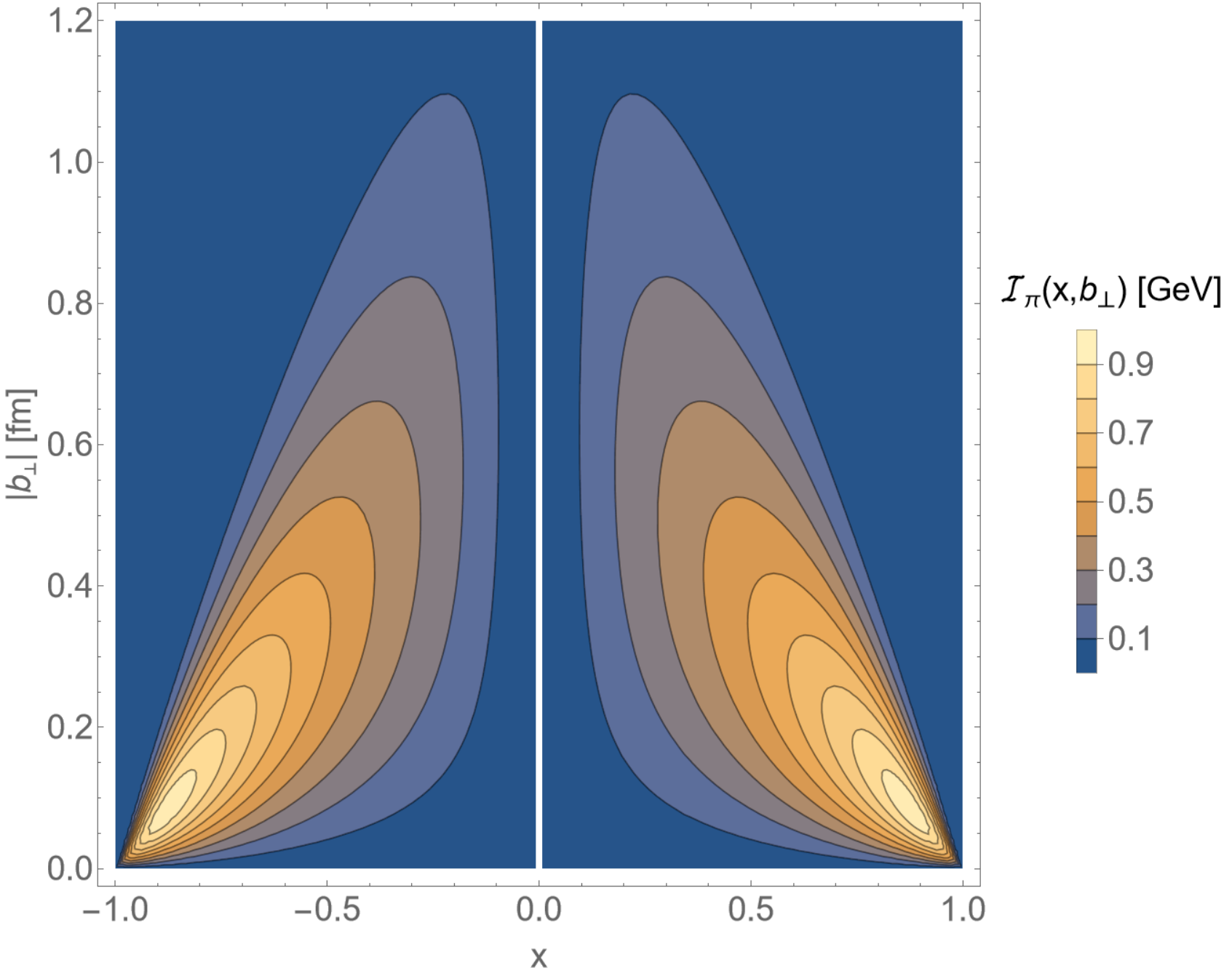}\hspace{0.3cm}\textbf{A} & \,\includegraphics[width=0.35\textwidth]{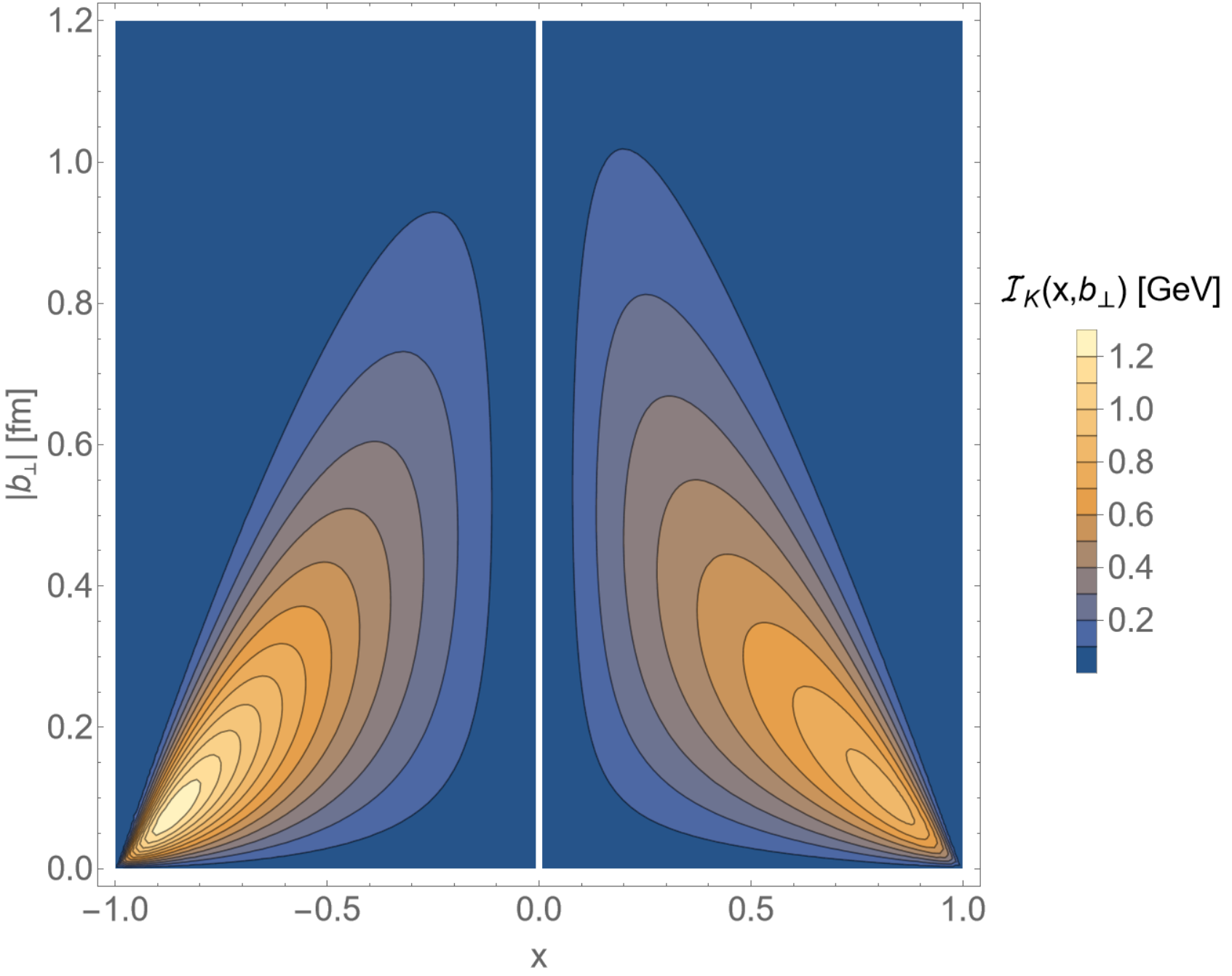}\hspace{0.3cm}\textbf{B} \\
\includegraphics[width=0.35\textwidth]{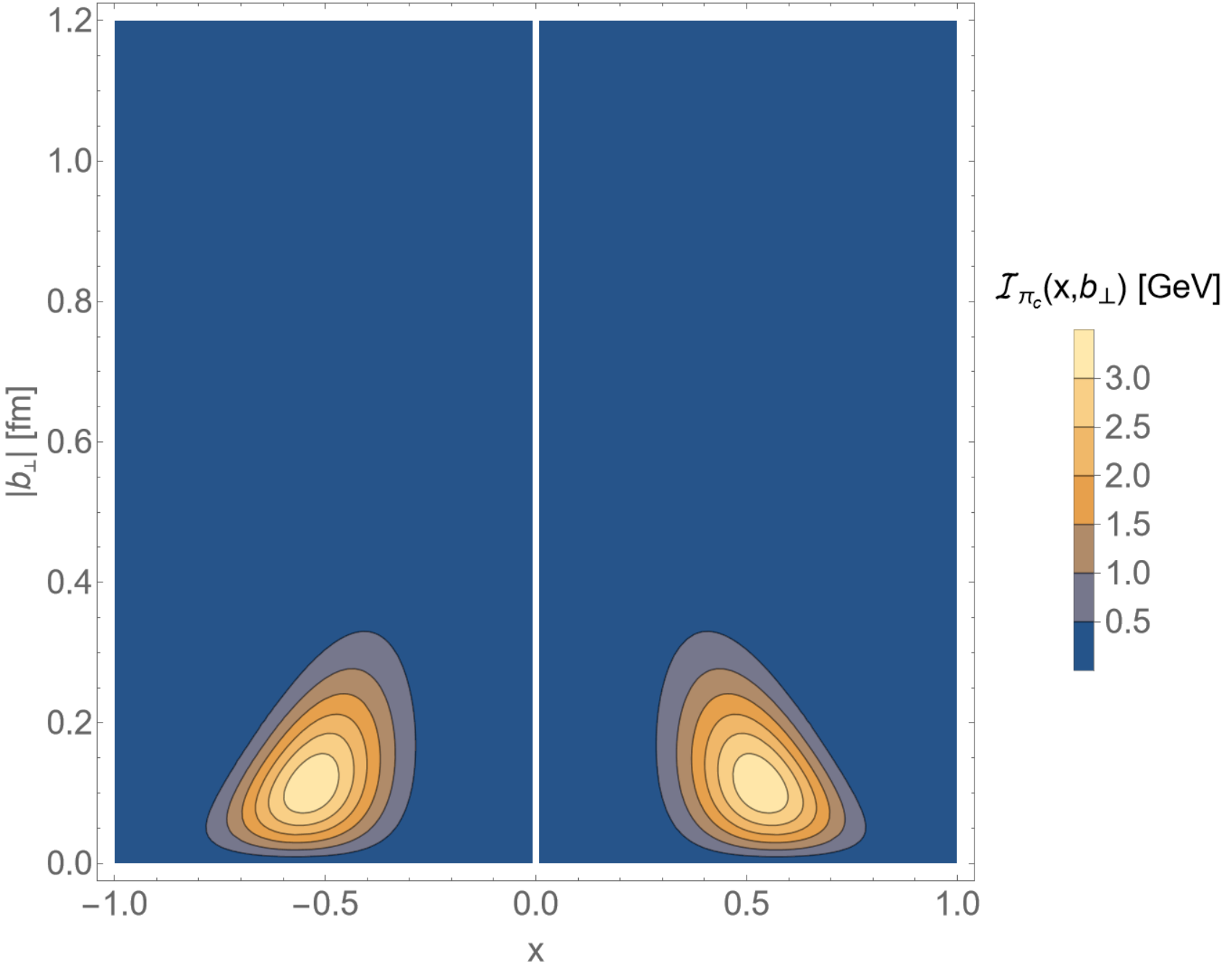}\hspace{0.3cm}\textbf{C} & \,\includegraphics[width=0.35\textwidth]{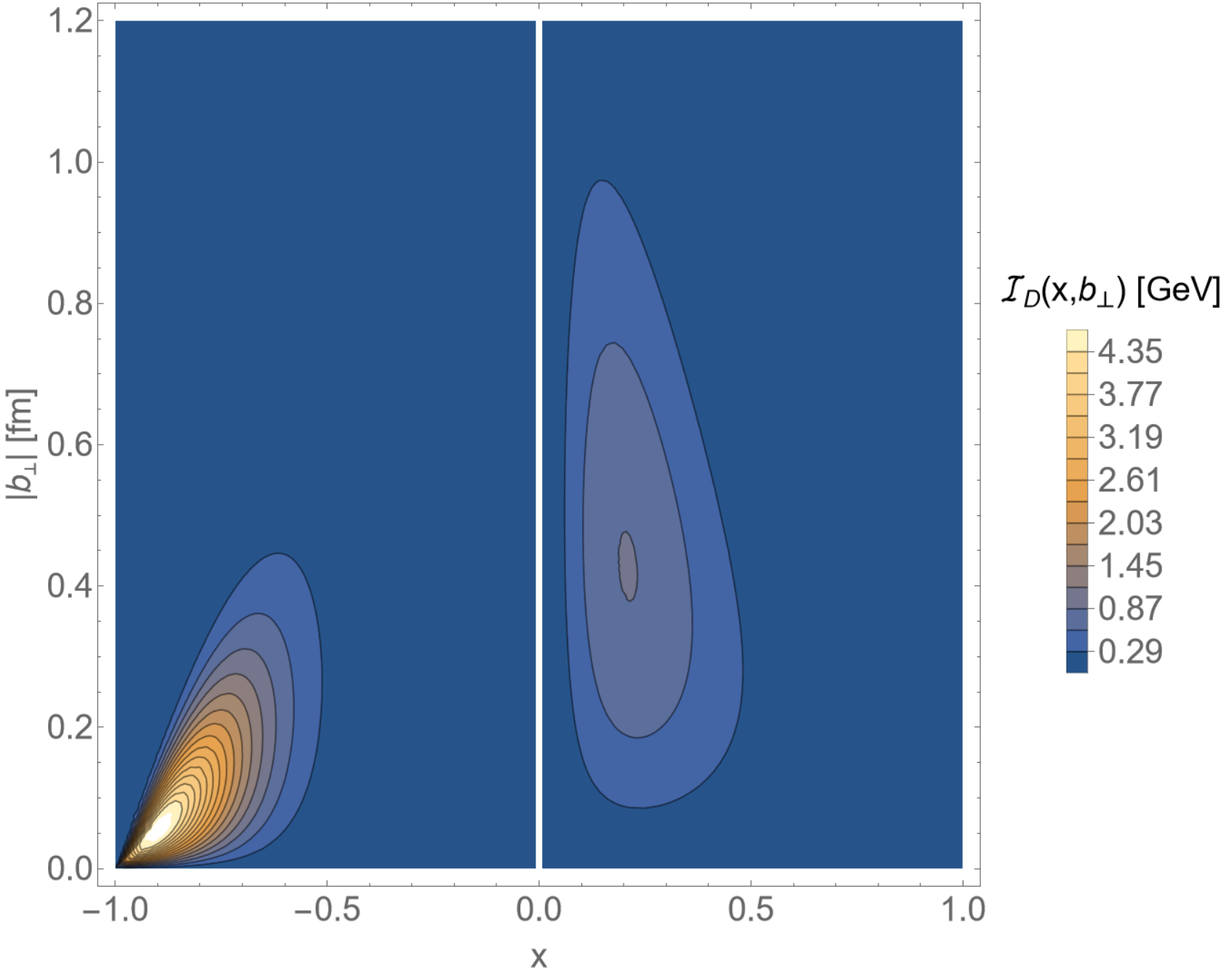}\hspace{0.3cm}\textbf{D}
\end{tabular}}
\caption{IPS-GPDs as defined in Eq.\,\eqref{eq:IPSdef}, evaluated at $\zeta_H$. The inputs are detailed in Section\,\ref{sec:numericalRes}.
[\textbf{A}] $\pi$. [\textbf{B}] $K^+$. [\textbf{C}] $\pi_c/\eta_c$. [\textbf{D}] $D^-=\bar{c}d$ meson.}
\label{fig:IPS}     
\end{figure*}

A detailed understanding of the spatial distribution within hadrons can be achieved through the IPS-GPDs, $\mathcal{I}_{\textbf{P}}^q(x,|b_\perp|;\zeta_H)$, which is defined in Eq.\,\eqref{eq:IPSdef}. Corresponding results for the systems being analyzed, specifically the $\pi,\,K,\,\pi_c$ and $D$ mesons, are displayed in Fig.\,\ref{fig:IPS}. Note that the distributions are obtained at $\zeta_H$ and the $x < 0$ domain corresponds to the antiquark. It is evident that in each instance, the IPS-GPD is maximal at a certain value of $(|x|^\text{max}, |b_\perp|^{\text{max}})$. The position and magnitude of this point is 
strongly correlated with the mass of the bound-state and its valence components. These metrics which are collected in Table\,\ref{tab:ImpParMax} and discussed below, happen to be compatible with previous explorations\,\cite{Zhang:2021mtn,Raya:2021zrz,Albino:2022gzs,Almeida-Zamora:2023bqb}.

\begin{table}[ht]
\centering
\caption{Global maximum of the IPS-GPD $\mathcal{I}_{\textbf{P}}^q(x,|b_\perp|;\zeta_H)$, $\mathscr{H}_{\textbf{P}}^q:=\text{max}\,\mathcal{I}_{\textbf{P}}^q(x,|b_\perp|)$, and its location. The IPS-GPD is obtained from Eq.\,\eqref{eq:IPSdef}, using the inputs described in Section\,\ref{sec:numericalRes}.}
\label{tab:ImpParMax}
\begin{tabular}[t]{l||c|c}
\hline
$\textbf{P}$ & $\{x^{\text{max}},\,|b_\perp|^{\text{max}}/\text{fm},\mathscr{H}\}_{\textbf{P}}^q\;$ & $\;\{-x^{\text{max}},\,|b_\perp|^{\text{max}}/\text{fm},\mathscr{H}\}_{\textbf{P}}^{\bar{h}}\;$ \\
\hline
$\pi$ & $0.88(2), 0.08(1),0.96(3)$ & $0.88(2), 0.08(1),0.96(3)$ \\
$K^+$   & $0.82(2), 0.10(1),0.85(3)$ & $0.86(2), 0.08(1),1.28(3)$  \\
$\pi_c/\eta_c$ & $0.53(2), 0.13(1),2.96(12)$ & $0.53(2), 0.13(1),2.96(12)$  \\
$D^-$ & $0.21(2), 0.42(2),0.90(4)$ & $0.91(2), 0.04(1),6.77(31)$  \\
\hline
\end{tabular}
\end{table}%

Capitalizing on the $\pi-K$ mesons, in the upper panels of Fig.\,\ref{fig:IPS}, the $\mathcal{I}_{\textbf{P}}^q(x,|b_\perp|;\zeta_H)$ distributions lean towards $|x| \to 1$ and show rather broad profiles. For the kaon, the slight asymmetry between the quark and antiquark distributions is nothing more than a reflection of the different constituent quark masses, which is of the order of $M_s/M_{u/d}\approx f_K/f_\pi\approx 1.2$. This implies that the $s$-quark has a more significant impact on determining the center of transverse momentum, but still with a contribution comparable to the light quark. 

  Returning to analyzing Fig.\,\ref{fig:IPS}, let's focus on other features regarding the $\pi_c$ and $D$ mesons depicted in the lower panels. Firstly, the IPS-GPD of the heavy-quarkonia system $\pi_c$ is significantly more compressed compared to its lighter counterparts, $\pi$ and $K$. In fact, the peak of the $\pi_c$ IPS-GPD is about 3 times higher than the pion's and tends to be located towards $|x|^{\text{max}}\to 1/2$, as would be expected for an infinitely massive system. On the other hand, the substantial asymmetry between the masses of the heavy and light quarks within the $D$ meson results in a distinctive profile of the IPS-GPD. Similar to the kaon, the heavier quark exhibits a narrower distribution, but has  a significantly higher maximum. Meanwhile, the light-quark profile is notably broad, with its peak lying within a low$-x$ and large$-|b_\perp|$ region. This indicates that the heavy quark carries most of the bound-state's total momentum, thereby defining the center of transverse momentum, with the light quark effectively orbiting around it. 

Such a perspective is further supported by the MSTE expectation values defined in Eq.\,\eqref{eq:MSTEave}, producing:
\begin{eqnarray}
    \langle |b_\perp|^2 \rangle^u_K = 0.71\, (r_E^K)^2 &,& \langle |b_\perp|^2 \rangle^{\bar{s}}_K = 0.59 \,(r_E^K)^2\,;\\
    \langle |b_\perp|^2 \rangle^d_D = 0.93 \,(r_E^D)^2 &,& \langle |b_\perp|^2 \rangle^{\bar{c}}_D = 0.14 \,(r_E^D)^2\,.
\end{eqnarray}

Regarding those systems in which flavor symmetry is assumed, such as the $\pi$ and $\pi_c$ states, a brief comparison shows that Eq.\,\eqref{eq:MSTEave} and Eq.\,\eqref{eq:radiiDEF}  lead to:
\begin{equation}
    \langle |b_\perp|^2 \rangle_\textbf{P}^{q,\bar{h}}=\frac{2}{3}(r_E^{\textbf{P}q,\bar{h}})^2=\frac{2}{3}(r_E^\textbf{P})^2\,.
    \label{eq:bperpiso}
\end{equation}
The first equality arises directly from the dimensionality of the corresponding expectation values, while the second is a consequence of flavor symmetry.

\section{Summary}
\label{sec:summary}
A variety of aspects regarding the internal structure of pseudoscalar mesons, particularly those revealed by the valence-quark GPD, have been addressed. For this purpose, we appeal to  its representation in Eq.\,\eqref{eq:GPDgen}, which we refer to as ERS and that relies on the prior knowledge of the corresponding DF and a given profile function that guides the off-forward behavior of the GPD. While the proposed parametrization is rather simple, and restricted to the zero-skewness limit,  it is quite general in its domain of applicability and enables the exploration of different features concerning the spatial distribution within the hadron, such as charge and mass distributions. Furthermore, it establishes relationships among various quantities, including electromagnetic and gravitational form factors, the IPS-GPD, as well as DAs and LFWFs. This allows the ERS to provide a comprehensive mapping of the internal structure of pseudoscalar mesons. More importantly, it generalizes and improves upon earlier efforts from the holographic approaches as well as the Gaussian and algebraic models proposed in the last years.

By primarily focusing on the NG modes, $\pi-K$, and their contrast with the $\pi_c/\eta_c$ and $D$ mesons—where the breaking of chiral and flavor symmetry is much more pronounced—the impact of the strong and weak mass generation mechanisms on the structural properties of hadrons is elucidated. Among other observations, a more significant $\Delta^2$-damping in the form factors is noted for light systems, leading to mass and charge density profiles that span over a larger spatial extent. In each instance, the mass distribution turns out to be more compact than the charge distribution. In systems exhibiting flavor symmetry, the profiles become more similar as the mass of the constituent quarks increases. The analysis of the $K$ and $D$ mesons indicates that in systems with a significant flavor asymmetry (such as heavy-light mesons), the heaviest quark plays a crucial role in defining the center of transverse momentum, effectively shaping the charge and mass densities. Intuitively, it suggests a physical scenario where the light quark occupies a relatively large region, orbiting around the heavier quark. These observations are further substantiated by the detailed visualization produced by the IPS-GPDs.  It clearly exposes structural  modifications brought about by a variation in the total mass of the system and its constituents. The inclusion of the artificial pseudoscalar $\pi_s$ enriches the physical intuition, as it demonstrates the traits of a composite state in which mass generation from QCD and the Higgs boson hold comparable dominance.  In this connection, the exploration of the special cases in the earlier sections enabled us to anticipate the trends in certain limits.

Notably, where comparisons are possible, the results generated by the ERS exhibit remarkable agreement with both experimental data and lattice QCD results, as well as with other theoretical approaches. By also considering the evolution of the collection of hadron distributions from a non-perturbative scale to the one accessible in experiment, further predictions can be made. This will be explored in future studies, following the all-orders approach outlined \emph{e.g.} in Refs.\,\cite{Cui:2020tdf,Raya:2021zrz,Yin:2023dbw,Lu:2023yna}. A more comprehensive effort that includes the remaining ground-state pseudoscalars is also in progress.

\section{Acknowledgment}
This work is supported by the Spanish MICINN grant PID2022-140440NB-C22, and the
regional Andalusian project P18-FR-5057.
A.~B wishes to acknowledge the {\em Coordinaci\'on de la Investigaci\'on Cient\'ifica} of the{\em Universidad Michoacana de San Nicol\'as de Hidalgo}, Morelia, Mexico,  grant no. 4.10, the {\em Consejo Nacional de Humanidades, Ciencias y Tecnolog\'ias}, Mexico, project CBF2023-2024-3544
as well as the Beatriz-Galindo support during his current scientific stay at the University of Huelva, Huelva, Spain. J. R-Q is also indebted to the \emph{Chair d'excellence} within the program \emph{d'Alembert} supporting a visiting professorship in the Universit\'e de Paris-Saclay, France. KR is grateful to Lei Chang for his valuable scientific insights.

\bibliography{main}
\end{document}